\documentclass[usenatbib]{mnras}

\usepackage[T1]{fontenc}
\usepackage{ae,aecompl}

\usepackage{verbatim} 
\usepackage{lipsum}
\usepackage{xcolor}
\usepackage{subfig}

\usepackage{graphicx}	
\usepackage{amsmath}	
\usepackage{amssymb}	

\title[IMBH random motion in star clusters]{Wandering off the centre: a characterization of the random motion of intermediate-mass black holes in star clusters}

\author[de Vita et al.]{
Ruggero de Vita$^{1}$\thanks{E-mail: r.devita@student.unimelb.edu.au},
Michele Trenti$^{1}$
and Morgan MacLeod$^{2, 3}$.
\\
$^{1}$School of Physics, The University of Melbourne, VIC 3010, Australia\\
$^{2}$Harvard-Smithsonian Center for Astrophysics, 60 Garden Street, Cambridge, MA 02138, USA\\
$^{3}$NASA Einstein Fellow\\
}

\date{Accepted XXX. Received YYY; in original form ZZZ}

\pubyear{2017}

\begin{document}
\label{firstpage}
\pagerange{\pageref{firstpage}--\pageref{lastpage}}
\maketitle

\begin{abstract}

Despite recent observational efforts, unequivocal signs for the presence of intermediate-mass black holes (IMBHs) in globular clusters (GCs) have not been found yet.
Especially when the presence of IMBHs is constrained through dynamical modeling of stellar kinematics, it is fundamental to account for the displacement that the IMBH might have with respect to the GC centre. In this paper we analyse the IMBH wandering around the stellar density centre using a set of realistic direct N-body simulations of star cluster evolution. Guided by the simulation results, we develop a basic yet accurate model that can be used to estimate the average IMBH radial displacement ($\left<r_\mathrm{bh}\right>$) in terms of structural quantities as the core radius ($r_\mathrm{c}$), mass ($M_\mathrm{c}$), and velocity dispersion ($\sigma_\mathrm{c}$), in addition to the average stellar mass ($m_\mathrm{c}$) and the IMBH mass ($M_\mathrm{bh}$). The model can be expressed by the equation $\left<r_\mathrm{bh}\right>/r_\mathrm{c}=A(m_\mathrm{c}/M_\mathrm{bh})^\alpha[\sigma_\mathrm{c}^2r_\mathrm{c}/(GM_\mathrm{c})]^\beta$, in which the free parameters $A,\alpha,\beta$ are calculated through comparison with the numerical results on the IMBH displacement. The model is then applied to Galactic GCs, finding that for an IMBH mass equal to 0.1\% of the GC mass, the typical expected displacement of a putative IMBH is around $1\arcsec$ for most Galactic GCs, but IMBHs can wander to larger angular distances in some objects, including a prediction of a $2.5\arcsec$ displacement for NGC 5139 ($\omega$ Cen), and $>10\arcsec$ for NGC5053, NGC6366 and ARP2. 

\end{abstract}



\section{Introduction} \label{sec:intro}

%

Investigating the existence of intermediate-mass black holes (IMBHs) in the Universe is a central goal in modern theories of galaxy evolution. In fact, IMBHs with masses in the range $10^2-10^5~M_\odot$ would represent the missing link between the well-known populations of the stellar BHs ($\lesssim 50~M_\odot$; see, e.g., \citealt{gies:86}, \citealt{orosz:11}, and the recent gravitational waves detections \citealt{abbott:16a}, \citealt{abbott:16b}, \citealt{abbott:17}), which represent the final result in the evolution of massive stars, and the supermassive BHs ($\gtrsim10^5~M_\odot$; see, e.g., \citealt{ghez:08}, \citealt{bentz:14}, \citealt{czerny:16}), which are ubiquitously observed at the centres of galaxies.

In recent years, globular clusters (GCs) have been indicated as promising candidates for hosting a central IMBH. Because of their high density cores ($\sim10^{3-4}~M_\odot$ pc$^{-3}$), they might represent optimal environments to grow IMBHs through runaway collapse of massive stars at their formation (see, e.g., \citealt{portegies:02}, \citealt{portegies:04} \citealt{giersz:15}). Alternatively, ejecta from first-generation massive stars might collect in the core of GCs, where the high gas density can lead to substantial accretion onto an existing BH seed (see \citealt{vesperini:10}).
Another motivation to establish whether IMBHs are present in GCs is connected to the $M_\bullet-\sigma$ relation (see \citealt{ferrarese:00,gebhardt:00}) between the mass of the supermassive BH and the average velocity dispersion in the bulge of galaxies. Given the range of typical core velocity dispersions in GCs, the existence of IMBHs would imply that this relation continues to be valid also in the low-mass regime (see, e.g., \citealt{lutzgendorf:13b}).


Yet, despite significant recent efforts to confirm (or falsify) these arguments, a clear evidence for the presence of IMBHs in GCs is still missing. 
Many different observational techniques can be used to identify IMBHs (see \citealt{mezcua:17} for a review). One possible method relies on the detection of both X-ray and radio emission associated, respectively,  to the accretion flow of gas onto the BH and to the synchrotron radiation from the emitted jets \citep{strader:12}. In particular, X-ray and radio observations can be used together to discriminate an IMBH against other plausible emitters and, thus, to set quantitative constraints on the IMBH mass (see \citealt{merloni:03}; \citealt{falcke:04}). These observations, however, are complicated by the typical lack of gas in GCs (see, e.g, \citealt{farrell:12}; \citealt{mezcua:13}; \citealt{haggard:13}). One illustrative case is represented by the centre of the globular cluster G1 in M31 \citep{pooley:06}, for which no clear sign of radio activity associated to the X-ray source has been found (see \citealt{ulvestad:07} and \citealt{miller-jones:12} for details). 

Another method for IMBH detection in GCs is represented by stellar dynamics measurements. This approach has led to the majority of the recent observational claims of detection in GCs, and, in a few instances, to debated results when applied to the same object (e.g., Omega Centauri in \citealt{noyola:10} and \citealt{vandermarel:10} or NGC 6388 in \citealt{lanzoni:13} and \citealt{lutzgendorf:13}, \citeyear{lutzgendorf:15}). For this method the constrains on the IMBH mass are generally determined by fitting the observed velocity dispersion profile with a family of Jeans models (see, e.g., \citealt{vandermarel:10}). These models are typically constructed by making assumptions on the mass-to-light ratio profile $M/L(r)$ in order to calculate the intrinsic mass distribution of the visible stars from the surface brightness profile. This represents a crucial and delicate step for the traditional Jeans modeling. In fact, a population of centrally concentrated dark remnants would be able to increase the inner $M/L$ profile and, thus, to produce effects on the kinematics of the luminous component similar to those of a central IMBH (see e.g., \citealt{arcasedda:16}; \citealt{peuten:17}; \citealt{gieles:17}). Further challenges for this method are represented by the necessity to make assumptions on the presence (or absence) of velocity anisotropy (see e.g. \citealt{zocchi:17}), and on the symmetry of the system. Finally, observations need to be able to measure accurately the velocity dispersion profile within the BH sphere of influence, which is expected to be limited to a few arcsec for most GCs. 

Another opportunity for IMBH detection is through the use of modern interferometers (see, e.g., \citealt{mandel:08}; \citealt{konstantinidis:13}; \citealt{macleod:16}), which can be able to detect the gravitational waves produced by possible mergers of massive compact objects into the IMBH. 

Finally, other complementary approaches are focused on indirect evidence of the IMBH presence. N-body numerical simulations suggest that a central IMBH is expected to form a shallow cusp in the projected surface brightness profile and to prevent the core collapse by enhancing three-body interactions within its sphere of influence (see e.g, \citealt{baumgardt:05}). In addition, IMBHs are shown to quench the phenomenon of mass segregation (see e.g. \citealt{trenti:07}; \citealt{gill:08}; \citealt{pasquato:16}) and energy equipartition \citep{trenti:13}, and such suppression may be constrained observationally through measurements of pulsar acceleration \citep{kiziltan:17}. However, measuring these effects in real GCs do not represent a sufficient condition to infer the existence of IMBHs, as other dynamical processes could be responsible for the same signatures (see e.g. \citealt{hurley:07}; \citealt{trenti:10}; \citealt{vesperini:10}).

Especially for dynamical modeling that rests on (spherical) symmetry assumptions, such as Jeans modeling of the surface brightness and velocity dispersion profiles, one possible source of systematic uncertainties may be represented by the wandering of the IMBH around the centre of the system (see e.g., \citealt{giersz:15}, \citealt{haster:16}, \citealt{devita:17}). 
From a dynamical point of view, the IMBH describes a random motion as a consequence of continual fluctuations in the global gravitational field induced by star encounters. The classical treatment for a point mass $M_\mathrm{bh}$, assuming energy equipartition with the background stars in the globular cluster's core, requires that $M_\mathrm{bh}\sigma_\mathrm{bh}^2=m_\mathrm{c}\sigma_\mathrm{c}^2$, where $m_\mathrm{c}$ is the typical mass of a field star and  $\sigma_\mathrm{bh}^2$, $\sigma_\mathrm{c}^2$ are the velocity dispersions of the IMBH and the field stars, respectively. Under the assumption that the IMBH moves in a specific gravitational potential, a simple model for the IMBH displacement can be obtained (see \citealt{bahcall:76}; \citealt{merritt:01}). However, this way of describing the IMBH motion is based on simplifying assumptions which need to be tested and potentially refined in order to reproduce the general behaviour found in N-body simulations.

The main goal of the present work is to produce a physically motivated model for the IMBH radial displacement by comparison with N-body simulations. The scope of this paper is dual. On one side we aim at identifying the main ingredients that contribute to the complex dynamics of IMBHs in star clusters. On the other side we aim at providing with a simple instrument to estimate the IMBH radial displacement on the base of few observational quantities. The model we propose represents an extension of the one discussed by \cite{bahcall:76}  (see Equation (\ref{eq:en_eq}) below), with two main physical ingredients added (degree of energy partition and core dynamical state), which will be constrained through comparison with numerical simulations. The paper is structured as follows. In Section \ref{sec:numres} we present the set of simulations used in this work together with important definitions for quantities relevant to our analysis. In Section \ref{sec:displ}, we derive a scaling relation which describes the IMBH average displacement in terms of relevant observational quantities by comparing our model to the simulations. In Section \ref{sec:real_clusters} we apply the model to an existing catalogue of 85 Galactic GCs in order to give reasonable predictions for the mean radial displacements of IMBHs that are assumed to constitute the 0.1\% of the total cluster mass. Finally, in Sect.\ref{sec:conclusions}, we give our conclusions.  

\section{Numerical framework} \label{sec:numres}

\subsection{Set of simulations} \label{subsec:simul}
The numerical simulations used in this paper are those from \cite{macleod:16}. 
The reader is directed to Section 2 of that paper for details. Here we summarise their main characteristics. 

The set of direct N-body simulations is produced by means of the NBODY6 distribution (\citealt{aarseth:99}, \citeyear{aarseth:03}) that embeds the SSE and BSE codes of \cite{hurley:00, hurley:02} to account for stellar evolution. Star clusters with $1-2\times10^5$ initial stars (corresponding to low-mass Galactic GCs) are evolved  in a realistic tidal field. The stellar distribution of each cluster is initialised following \citet{king:66} models with $W_0=7$. The stars in the initial conditions follow a \citet{kroupa:01} initial mass function (IMF), within the mass range $0.1-30~M_\odot$, with no binaries. The metallicity is one-tenth solar.

At the beginning of the simulation, an IMBH of $75-150~M_\odot$ and with zero velocity is initialised at the centre of mass of the system. The IMBH mass grows modestly during the evolution because of tidal disruption events following close encounters (typically, by the end of the evolution, the IMBH mass increases by a factor $1.2-1.4$, depending on the simulation group). In these events, which take place when the pericenter radius is less than $r_\mathrm{t}=(M_\mathrm{bh}/M_*)^{1/3}R_*$, a fraction of the star mass is accreted into the IMBH (for weakly-bound orbits typically half of the mass is retained by the BH). 

Finally, \citet{macleod:16} consider two different cases for the velocity kicks imparted to stellar remnant (neutron stars and BHs). In both cases remnants are given a kick drawn from a Maxwellian distribution with sigma of either 1 or 2.5 times the initial cluster velocity dispersion, producing different retention fractions of stellar remnants.

The simulations have been divided in 4 groups, which differ for the initial parameters. Within each group, statistically different realisations of the same initial conditions are considered. The main properties of the simulations groups are summarised in Table \ref{tab:simul}.

 \begin{table}
\centering
\caption{Table of N-body simulation groups A-D. For each group we report (from left to right) the number of initial stars ($N_*$); the number of equivalent simulations ($N_\mathrm{sim}$), which are different realisations of the same initial conditions; the King parameter $W_0$ and the initial half-mass radius $r_{\mathrm{h},0}$ in pc; the initial IMBH mass in $M_\odot$; the total duration of the simulation in Gyr, and the kick imparted to stellar remnants in terms of the initial cluster velocity dispersion $\sigma_*$.} 
\label{tab:simul}
\begin{tabular}{cccccccc}
\hline
 & $N_*$ & $N_\mathrm{sim}$ & $W_0$ &$r_{\mathrm{h},0}$ &$M_{\mathrm{bh},0}$ &$t_\mathrm{max}$ &$\sigma_k/\sigma_*$   \\
\hline

A 	& 100k	&3 	& 7 	& 2.3 	&150 	& 6 		& 2.5\\
B 	& 100k 	&3	& 7 	& 2.3 	&150 	& 6 		& 1.0\\
C 	& 100k 	&3	& 7 	& 2.3 	& 75  	& 9 		& 2.5\\
D 	& 200k 	&4	& 7 	& 2.3 	&150 	& 10.4 	& 2.5\\
\hline
\end{tabular}
 \end{table}
 
\subsection{IMBH displacement definition} \label{subsec:quant}
In order to characterise the motion of the IMBH in our simulated GCs with respect to the GC centre, we need to properly define a coordinate-independent centre of the system. Following what suggested by \citet{casertano:85}, we use a density centre for the purpose. For this, we associate to each particle in a snapshot (excluding the IMBH) a local density calculated considering the six closest neighbours. In particular,  we define the local density for the i-th star as
\begin{equation}
\rho_i=\frac{M-m_i-m_6}{4/3\pi r_6^3},
\end{equation}
where $m_i$ is the mass of the i-th star, $m_6$ and $r_6$ are the mass and the distance of the sixth neighbour to the i-th star, respectively, and $M$ is the total mass within a sphere of radius $r_6$ centred in the i-th star position.
The density-weighted centre of the system $\vec{x}_{\rho}$ is then defined by
\begin{equation}
\label{xdc}
\vec{x}_{\rho}=\frac{\sum_i\vec{x}_i\rho_i}{\sum_i\rho_i},
\end{equation}
where $\vec{x}_i=(x_i,y_i,z_i)$ is the position of the i-th star with respect to the initial reference frame.
Finally, we consider a mass/density-weighted radius defined as
\begin{equation}
\label{rd}
r_{\rho}=\frac{\sum_i\left|\vec{x}_i-\vec{x}_{\rho}\right|\rho_i m_i}{\sum_i\rho_i m_i}.
\end{equation}
As tested in \citet{trenti:10}, the density radius $r_\rho$ generally represents a good estimate for the core radius as determined by fitting the surface-brightness profile with a King model, especially for systems that are not core-collapsed (see Fig.~10 in \citealt{trenti:10}). 
With these definitions, the IMBH radial position can be expressed in a coordinate-independent way as
\begin{equation}
r_\mathrm{bh} = \sqrt{|\vec{x}_\mathrm{bh}-\vec{x}_\rho|^2}.
\end{equation}

All the definitions given so far can be easily extended in order to deal with projected and luminosity-weighted quantities. We define the local surface-brightness for the i-th star as
\begin{equation}
\mu_i=\frac{L-l_i-l_6}{\pi R_6^2},
\end{equation}
where $l_i$ is the luminosity of the i-th star, $l_6$ and $R_6$ are the luminosity and the projected distance of the sixth neighbour to the i-th star, respectively, and L is the total luminosity within a circle of radius $R_6$ centred in the i-th star position. We calculate luminosity-weighted quantities by considering only main-sequence stars, thus avoiding the fluctuations in the light profile that arise from the small number of luminous giants. Moreover, we consider stars with mass higher than $0.4M_\odot$ in order to exclude the faint end of the main sequence, since those stars would be likely unresolved in typical GC observations. For main sequence stars we assume that luminosity scales with mass as $l\propto m^{7/2}$.
Then, the 2 dimensional luminosity centre $\vec{X}_{\mu}$ is obtained by projecting the stars' positions along the z-axis and by replacing $\rho_i$ with $\mu_i$ in Equation (\ref{xdc}):
\begin{equation}
\label{Xmu}
\vec{X}_{\mu}=\frac{\sum_i\vec{X}_i\mu_i}{\sum_i\mu_i},
\end{equation}
where $\vec{X}_i=(x_i,y_i)$. Finally, analogous to the mass/density-weighted radius in Equation (\ref{rd}) is the surface-brightness density radius $R_\mu$, which is defined by
\begin{equation}
\label{Rmu}
R_{\mu}=\frac{\sum_i\left|\vec{X}_i-\vec{X}_{\mu}\right|\mu_i l_i}{\sum_i\mu_i l_i}.
\end{equation}

\subsection{IMBH displacement results} \label{subsec:displ_sim}

\begin{figure*}
{\includegraphics[width=1.0\textwidth]{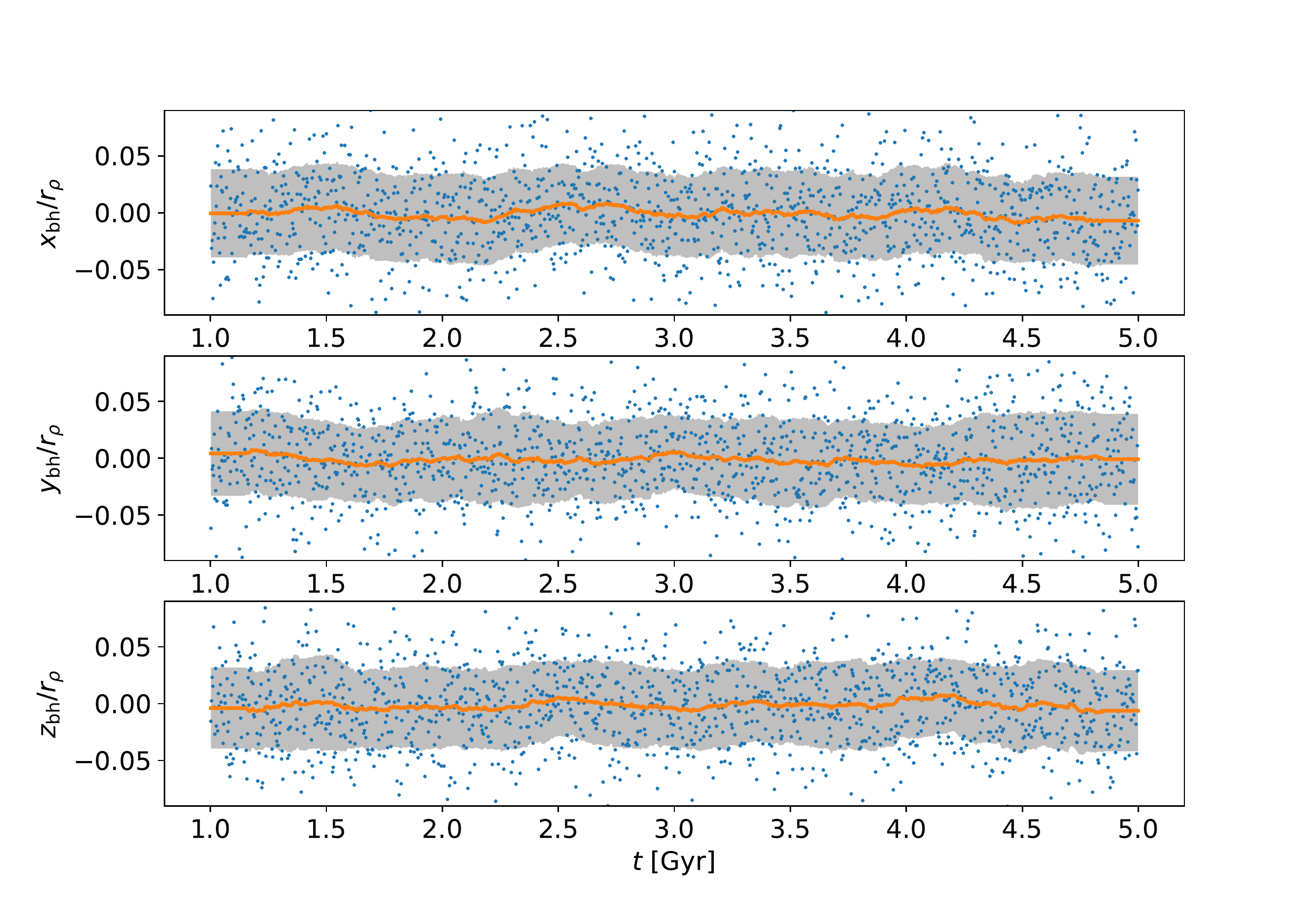}}
\caption{Evolution of the IMBH displacement relative to the density radius ($\sim$ core radius) for one simulation in group A in each direction. Each data point represents a single measurement coming from one snapshot of the simulation. The orange line is a running average, calculated by averaging the values obtained from 50 snapshots, corresponding to roughly 150 Myr. The shaded area encloses two standard deviations with respect to the running average (one above and one below).}
\label{fig:evol_displ_xj}
\end{figure*}

For one simulation in group A, we report in Fig.~\ref{fig:evol_displ_xj} the evolution of the ratio of the IMBH displacement along different axes as a function of time. We exclude the first part of the evolution from our analysis because, in this phase, the internal dynamics is still dominated by rapid stellar evolution processes. In particular, there is a  significant mass loss as higher mass stars evolve out of the main sequence. From the figure it is evident how the IMBH is experiencing an isotropic motion around the centre. Together with the single values from each snapshot in the simulation, we plot also a running average for the sample. It is worth noting that, when rescaled with the density radius, the displacement on each axis is approximately time-independent. 

\begin{figure*}
\begin{tabular}{cc}
\subfloat[]{\includegraphics[width=0.47\textwidth]{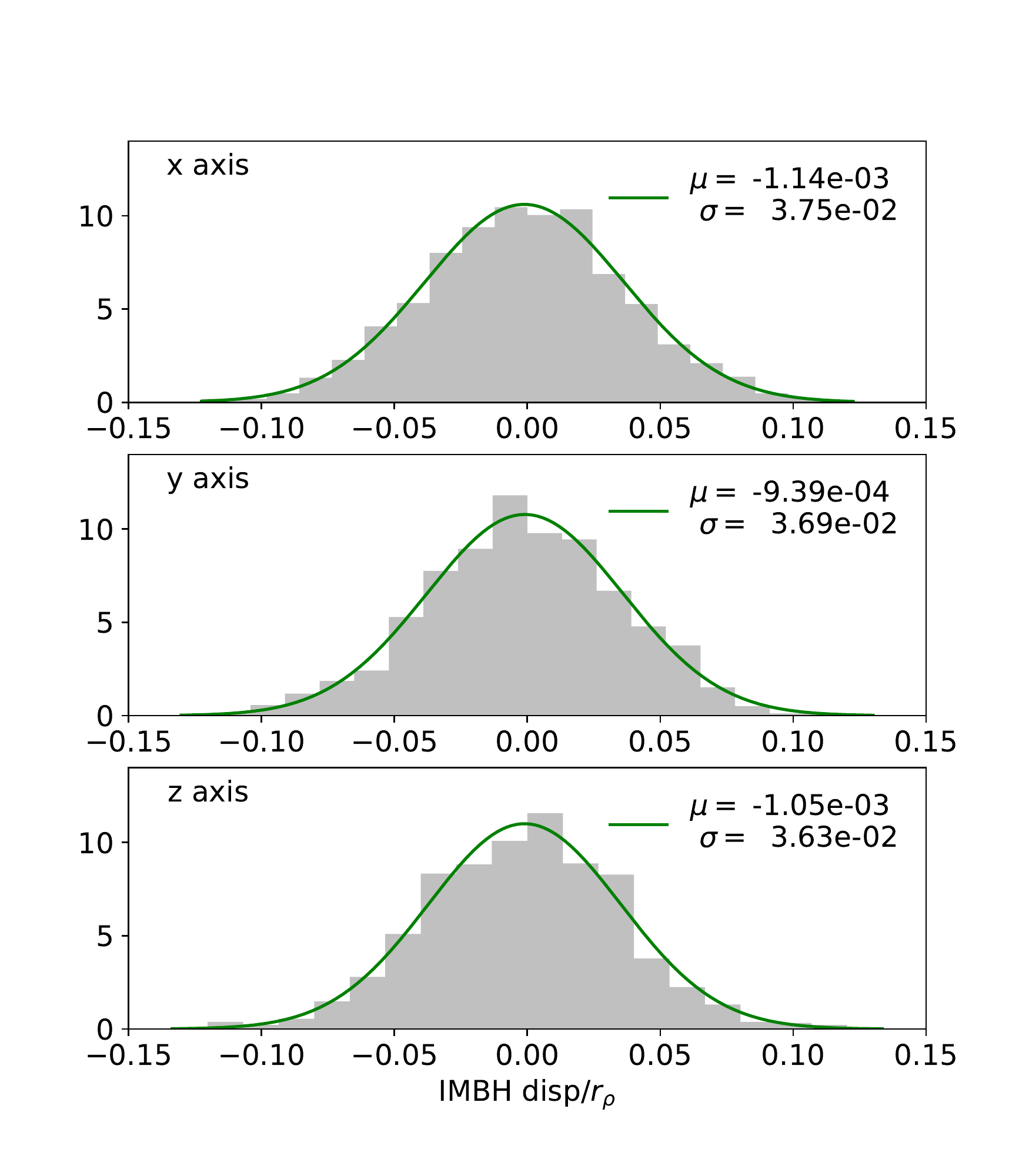}} &
\subfloat[]{\includegraphics[width=0.47\textwidth]{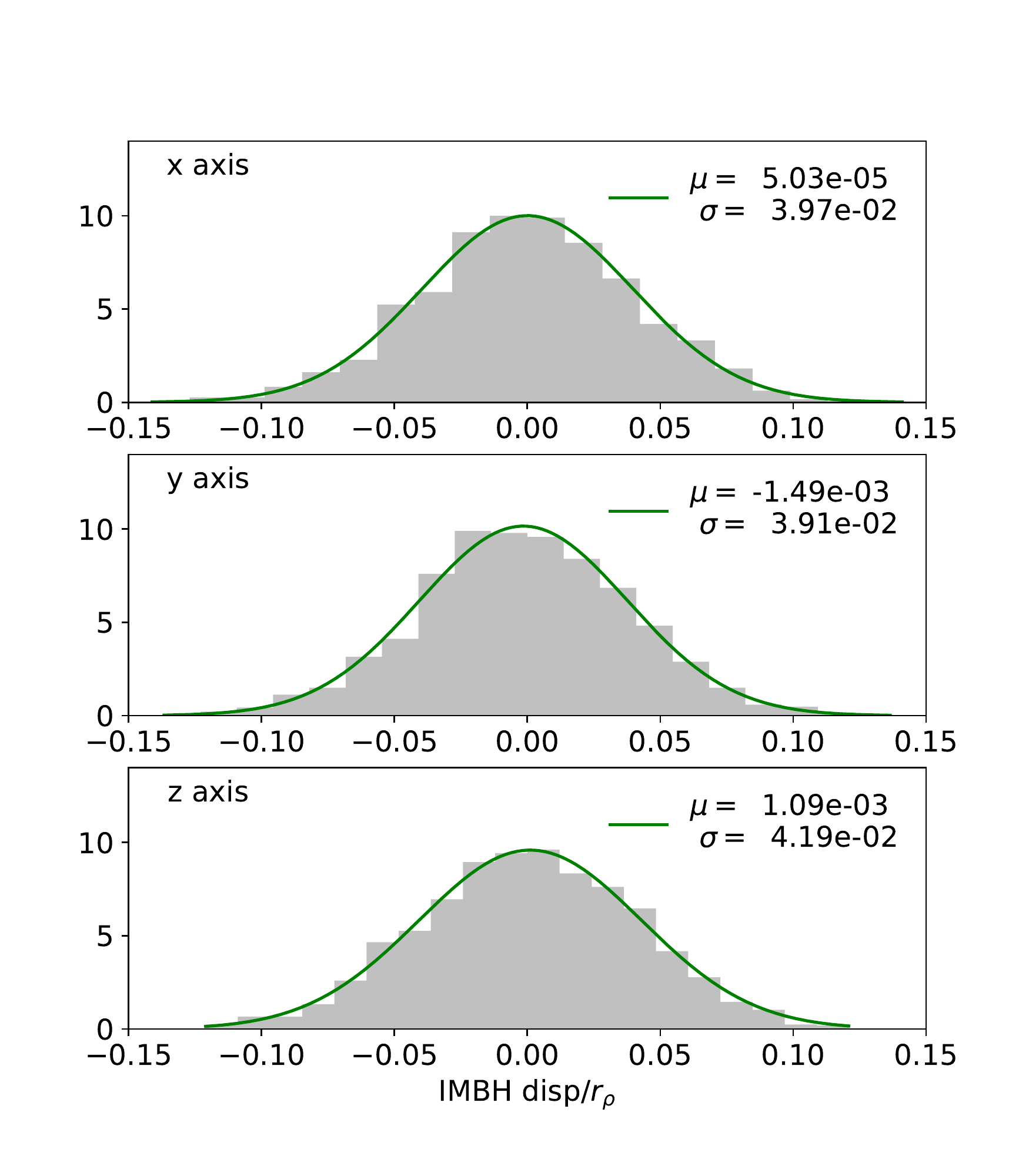}}
\\
\subfloat[]{\includegraphics[width=0.47\textwidth]{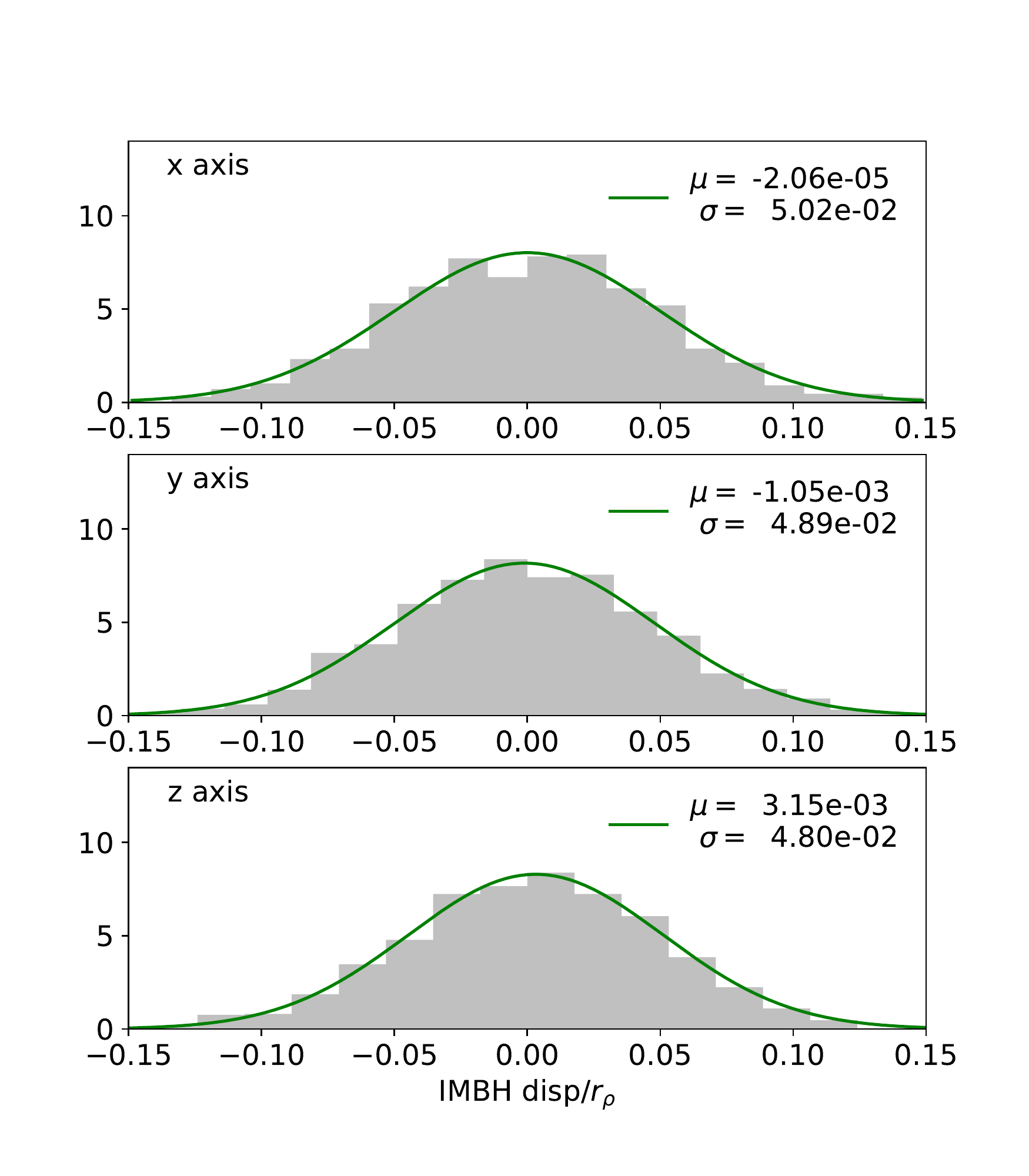}} &
\subfloat[]{\includegraphics[width=0.47\textwidth]{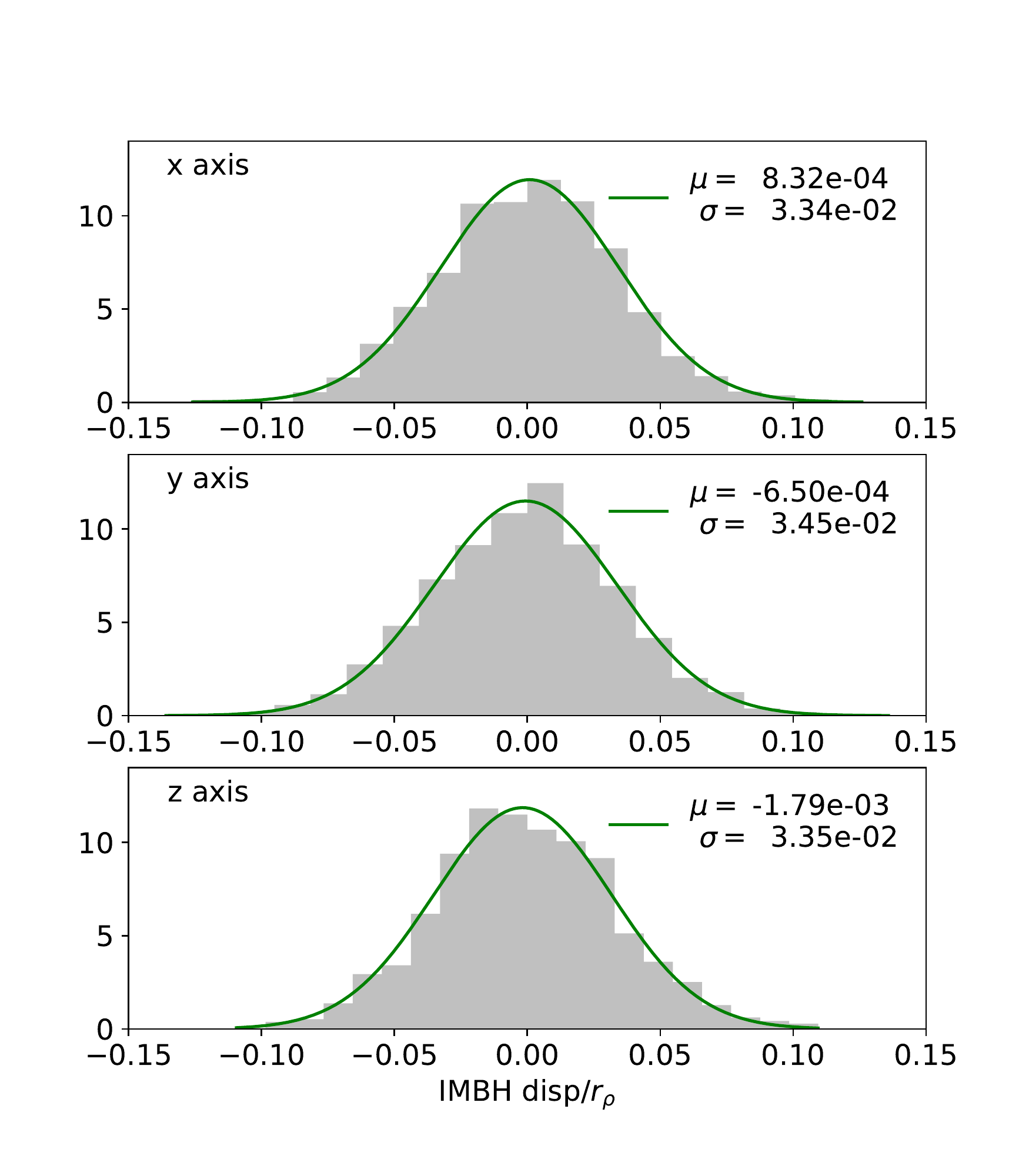}}
\\
\end{tabular}
\caption{Distribution of the ratio between the IMBH displacement and the density radius in three directions $(x, y, z)$ for the 4 groups of simulations used in this paper (see Table \ref{tab:simul}) . Each distribution refers to the time interval 1-5 Gyr and is fitted with a Gaussian distribution with mean value $\mu$ and standard deviation $\sigma$. The average displacement, identified as the $\sigma$ of the best-fit normal distribution, is higher in panel C which shows the simulation with the less massive IMBH (75 $M_\odot$).}
\label{fig:meas_displ}
\end{figure*}

In Fig.~\ref{fig:meas_displ} we show the displacement distribution in the range 1-5 Gyr for 4 typical simulations, one for each group analysed in this paper (see Table \ref{tab:simul}). As expected, the distributions in each direction are well fitted by a normal distribution with zero mean, suggesting that the IMBH is actually experiencing a Brownian random motion. This motivates and justifies the approach to consider only radial quantities for further analysis.

\begin{figure*}
{\includegraphics[width=1.0\textwidth]{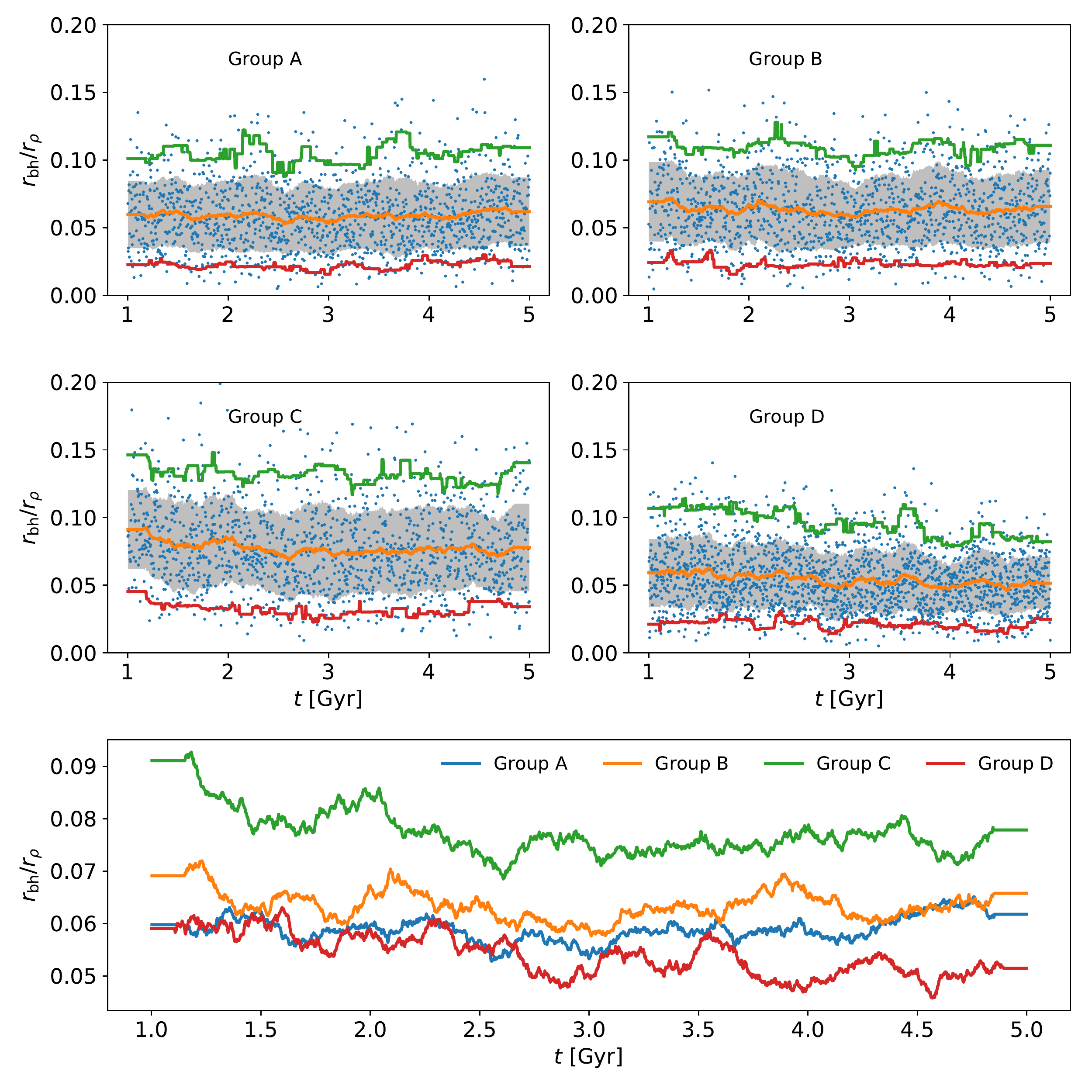}}
\caption{Evolution of the IMBH displacement for 4 simulations, one for each group analysed in this paper (see Table \ref{tab:simul}). {\it Upper panels: } each data point represents a single measurement of the ratio $r_\mathrm{bh}/r_\rho$ coming from one snapshot of the simulation. The orange, red and green line represent respectively the running average, the 5\% quantile and the 95\% quantile, calculated from the distribution of the values obtained from 50 snapshots, corresponding to roughly 150 Myr for groups A-B-C and to 100 Myr for group D. The shaded area encloses two standard deviations with respect to the running average (one above and one below). The average IMBH displacement slightly varies during the evolution as a direct consequence of processes as mass segregation in the core ($m_\rho$ increases), IMBH accretion ($M_\mathrm{bh}$ increases) and core collapse ($r_\rho$ decreases). {\it Lower panel:} Running averages for each group above but shown in a a single plot for direct comparison. On average, the displacement in group C (low-mass IMBH) is significantly higher than the others.}
\label{fig:evol_displ_r}
\end{figure*}

Finally, in Fig.~\ref{fig:evol_displ_r} we report the radial displacement evolution for the 4 groups of simulations considered. For all the simulations, the minimum average displacement for the IMBH is $\gtrsim0.02\ r_\rho$ with 95\% confidence. According to the classic derivations for the Brownian motion of a point mass object, the IMBH mean radial displacement $\left<r_\mathrm{bh}\right>$ in terms of the density radius $r_\rho$ is represented by
\begin{equation} \label{eq:en_eq}
\frac{\left<r_\mathrm{bh}\right>}{r_\rho}\approx\left(\frac{m_\rho}{M_\mathrm{bh}}\right)^{1/2},
\end{equation}
where $m_\rho$ is the average stellar mass within the density radius. This result is obtained by \citealt{bahcall:76} under the assumptions that the IMBH is a single object in complete energy equipartition with the surrounding stars in the core and is moving in an harmonic potential well. 
For typical values in our simulations, $m_\rho\approx0.65 M_\odot$ and $M_\mathrm{bh} \approx 150 M_\odot$, we have $\left<r_\mathrm{bh}\right>/r_\rho\approx0.07$,  which is in good agreement (at least as order of magnitude estimate) with the measured displacements. In addition, the radial IMBH displacement increases significantly in the group of simulations with the less massive IMBH (see group C). Moreover, both the different way of treating the velocity kicks imparted to stellar remnants and the initial number of stars seems to play a secondary role in the IMBH wandering (compare groups A, B and D).

\renewcommand{\thesubfigure}{\Alph{subfigure}}

\section{A scaling relation for the IMBH displacement} \label{sec:displ}

\subsection{Physical foundations}
\label{subsec:model}
The basic assumption behind Equation (\ref{eq:en_eq}) is that the IMBH has reached a state of energy equipartition with the surrounding stars in the core. As shown in \cite{trenti:13}, a condition of complete energy equipartition is not achieved in the context of direct N-body simulations. Thus, for our purposes, we will consider a more general relation between the IMBH velocity dispersion $\sigma_\mathrm{bh}$ and the core velocity dispersion $\sigma_\mathrm{c}$:
\begin{equation} \label{eq:part_en_eq}
\sigma_\mathrm{bh}^2 = \left(\frac{m_\mathrm{c}}{M_\mathrm{bh}}\right)^{2\alpha}\sigma_\mathrm{c}^2,
\end{equation}
where $\alpha$ is a free parameter that measures the degree of energy equipartition in the core (we expect $0\leq\alpha<0.5$). 

If we Taylor expand a King potential \citep{king:66} in $\left<r_\mathrm{bh}\right>\approx0$ and we equal the IMBH kinetic energy with its gravitation potential energy,  we are able to associate the IMBH velocity dispersion to a physical mean displacement. In particular, given the total mass $M_\mathrm{c}$ enclosed in the core radius $r_\mathrm{c}$, we have
\begin{equation} \label{eq:king_pot}
\sigma_\mathrm{bh}^2 \propto \frac{GM_\mathrm{c}}{r_\mathrm{c}} \left(\frac{\left<r_\mathrm{bh}\right>}{r_\mathrm{c}}\right)^2, 
\end{equation}
where, for King models, the constant of proportionality is determined by fixing the central dimensionless potential $W_0$. By substituting $\sigma^2_\mathrm{bh}$ using the combination of Equation (\ref{eq:part_en_eq}) and Equation (\ref{eq:king_pot}),   the mean IMBH displacement can be expressed as
\begin{equation} 
\frac{\left<r_\mathrm{bh}\right>}{r_\mathrm{c}} \propto \left(\frac{m_\mathrm{c}}{M_\mathrm{bh}}\right)^\alpha \left(\frac{\sigma_\mathrm{c}^2r_\mathrm{c}}{GM_\mathrm{c}}\right)^{1/2}.
\end{equation}
Finally, in order to be more general, we define the exponent of the factor $\sigma_\mathrm{c}^2r_\mathrm{c}/(GM_\mathrm{c})$  as $\beta$ and consider it as a free parameter related to the dynamical state of the core.

In conclusion, we obtain the following equation for the average IMBH displacement
\begin{equation} \label{eq:displ}
\frac{\left<r_\mathrm{bh}\right>}{r_\mathrm{c}} = A \left(\frac{m_\mathrm{c}}{M_\mathrm{bh}}\right)^\alpha \left(\frac{\sigma_\mathrm{c}^2r_\mathrm{c}}{GM_\mathrm{c}}\right)^\beta, 
\end{equation}
with $A$, $\alpha$ and $\beta$ free parameters of the model. 

The main goal of this work is to find the parameters' values that best represent our simulations. We wish to emphasis here that our model intentionally relies on a low number of free parameters. This choice is motivated by the aim of describing a complex dynamical phenomenon with the minimal number of ingredients, which are based on understandable and basic physical arguments. Despite its simplicity, the number of free parameters is still higher with respect to the model of Equation (\ref{eq:en_eq}) presented by \cite{bahcall:76}, but we show in Subsection \ref{subsec:mass_based_analysis} that the addition of $\beta$ is supported by data-model comparison.

\subsection{Binary versus three-body interactions}
\label{subsec:2bvs3b}

One possible extension for the physical treatment presented in the previous Subsection is to consider the IMBH as one component of a binary system (see \citealt{merritt:01}), which is the most likely IMBH dynamical state observed in numerical simulations (see \citealt{macleod:16}). The Brownian motion of a binary in a background field differs from that of a single massive object because of inelastic scattering events. These occur when a perturber star strongly interacts with the binary, and is ejected after one or several encounters carrying away part of the binary binding energy, thus not conserving the total kinetic energy of the three-body system. If the binary mass is much greater than the average field mass, the net result of many close three-body interactions is to increase the recoil velocity of the binary centre of mass as a consequence of linear momentum conservation. 

\citet{merritt:01} compares the rate of diffusion in the velocity of the binary due to three-body superelastic scatterings $\left<(\Delta v)^2\right>_\mathrm{se}$, with that of a point mass due to two-body encounters $\left<(\Delta v)^2\right>_\mathrm{C}$ (see Equations 3 to 11 in that paper). The ratio of these quantities is given by
\begin{equation}
\label{eq:3vs2}
\frac{\left<(\Delta v)^2\right>_\mathrm{se}}{\left<(\Delta v)^2\right>_\mathrm{C}}\sim \frac{H}{32\sqrt{2\pi}\log\Lambda},
\end{equation}
where $\log\Lambda$ is the Coulomb logarithm, which for typical GCs is $3\lesssim\log\Lambda\lesssim5$ (see, e.g., \citealt{bertin:14}) and $H$ represents the hardening rate of the binary. From three-body scattering experiments, even for hard binaries (i.e., those for which $\sqrt{GM_{12}/a}\gg \sigma_\mathrm{c}$, with $M_{12}$ total binary mass and $a$ semi-major axis), $H\lesssim20$ for a wide range of mass ratios (see \citealt{quinlan:96}). Thus, Equation (\ref{eq:3vs2}) implies that the enhancement of the IMBH random motion via three-body encounters might be negligible at first instance ($\approx6\%$ correction). This conclusion seems even more appropriate for massive star clusters (we recall that the numerical simulations used in this paper may only represent the low-mass end of the Galactic GCs' system). In fact, we expect the $\log\Lambda$ term in Equation (\ref{eq:3vs2}) to slightly increase with the number of stars, reducing the impact of three-body encounters on the binary diffusion rate. 

\begin{figure*}
\includegraphics[width=\textwidth]{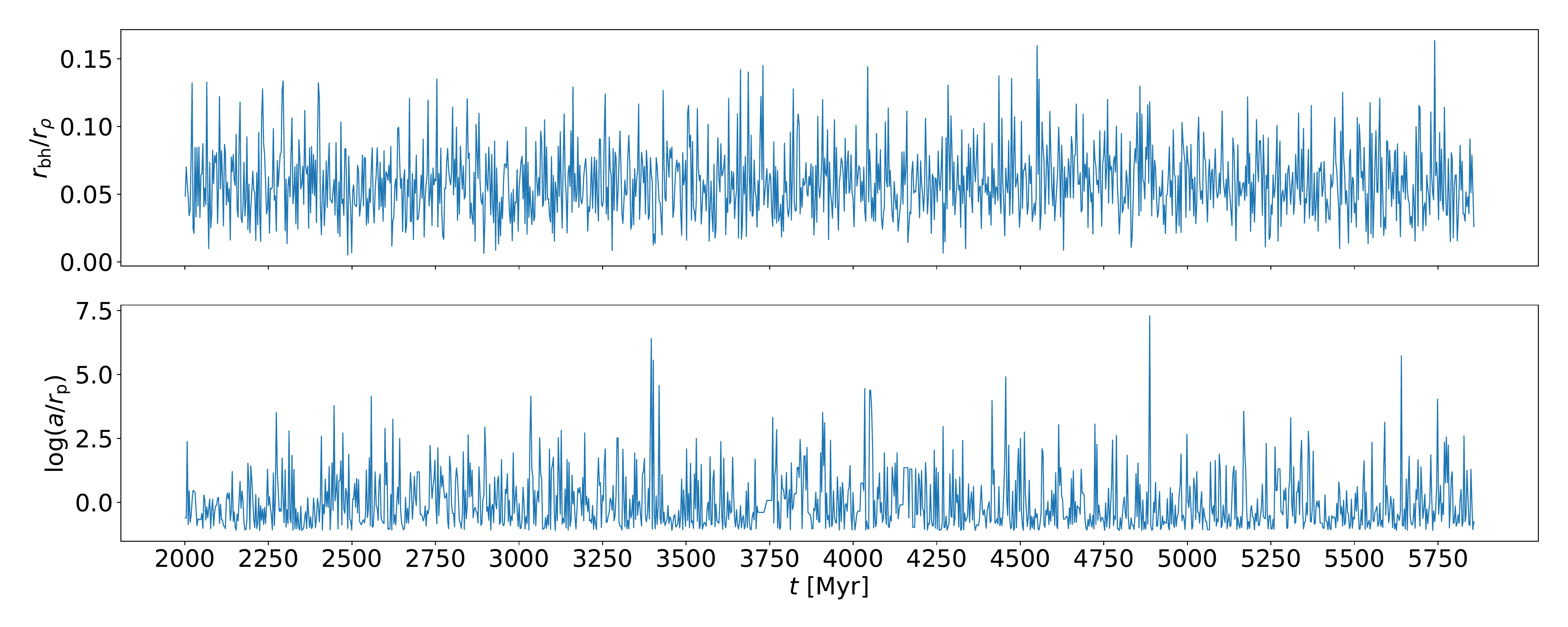}\\
\includegraphics[width=\textwidth]{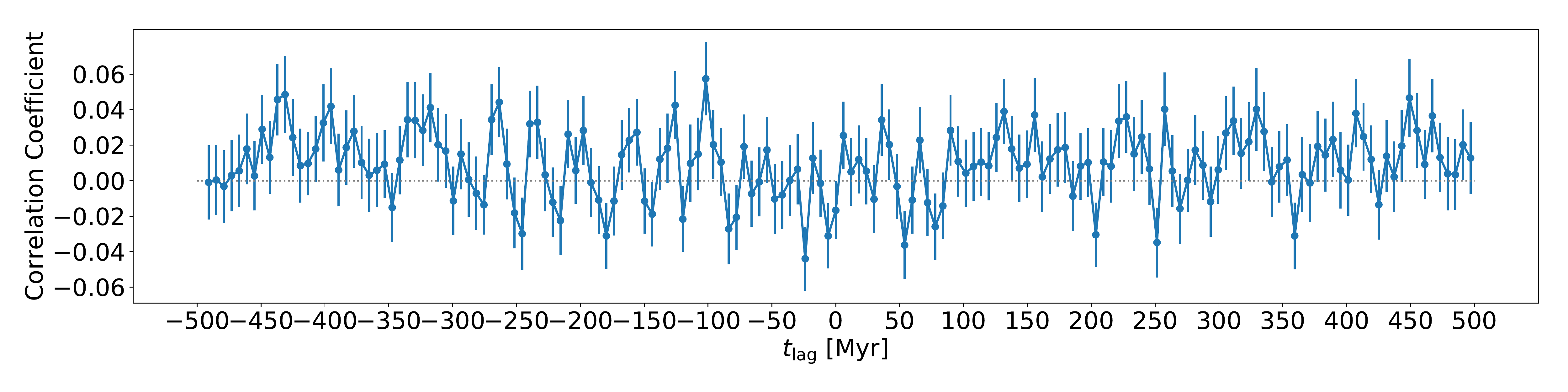}
\caption{Cross-correlation of the IMBH radial displacement with three-body interactions. {\it Upper panels:} evolution of the IMBH radial displacement compared to the most energetic three-body events quantified by the ratio $a/r_\mathrm{p}$ of the semi-major axis of a binary with the IMBH and the closest distance of a perturber star. {\it Lower panel:} discrete correlation coefficients of the two time signals $r_\mathrm{bh}/r_\rho$ and $a/r_\mathrm{p}$ as function of the lag time. The low and uniformly distributed values of the coefficients might suggest that the IMBH radial displacement is not significantly affected by strong three-body encounters. The dynamical time for the simulation is $t_\mathrm{dyn}\approx0.3$ Myr.}
\label{fig:bh_3bdy}
\end{figure*}

To further investigate the impact of three-body interactions, we directly searched for a correlation between the IMBH radial displacement and three-body scattering events in our fiducial simulation (group A in Table \ref{tab:simul}). We flagged the three-body interactions as those where the IMBH is one component of an hard binary system for which $r_\mathrm{p}/a < 3$, with $r_\mathrm{p}$ representing the closest approach distance of a perturber star to the binary. In Fig.~\ref{fig:bh_3bdy} we plot the displacement evolution in comparison with the ratio $a/r_\mathrm{p}$. In order to characterise a possible dependence between the two signals, we calculated the discrete correlation function by means of the \texttt{pyDCF} software (see \citealt{robertson:15}), which is specifically designed to deal with unevenly sampled time series. We expect this function to be peaked at a certain lag time $t_{\rm{lag}}$ if $r_\mathrm{bh}$ increments follow close three-body scattering events (those for which $a \gg r_\mathrm{p}$) with a characteristic time delay $t_{\rm{lag}}$. The cross-correlation coefficients are shown in Fig.~\ref{fig:bh_3bdy}. The time range considered for $t_{\rm{lag}}$ is of the order of the typical half-mass relaxation time $0.5 \rm{Gyr} \lesssim t_{\rm{rh}} \lesssim 1.5 \rm{Gyr}$ (see \citealt{trenti:10} for a definition). Note that we plot the amplitudes of the correlation also for negative values of $t_{\rm{lag}}$. Quantifying the amplitude of such correlations (which are not expected to be present in the system) provides us with a characteristic noise level that can be directly and easily compared with the amplitudes of interest in the range $t_{\rm{lag}}>0$, showing that there is no difference in amplitudes and thus that the correlations at positive lag times are dominated by noise. 

Overall, all these tests confirm that inelastic scattering events are not dominating the IMBH radial displacement variations and thus we neglect them in the following analysis. 


\subsection{Results}

\renewcommand{\thesubfigure}{\Alph{subfigure}}
\begin{figure*}
\begin{tabular}{cc}
\subfloat[]{\includegraphics[width=0.47\textwidth]{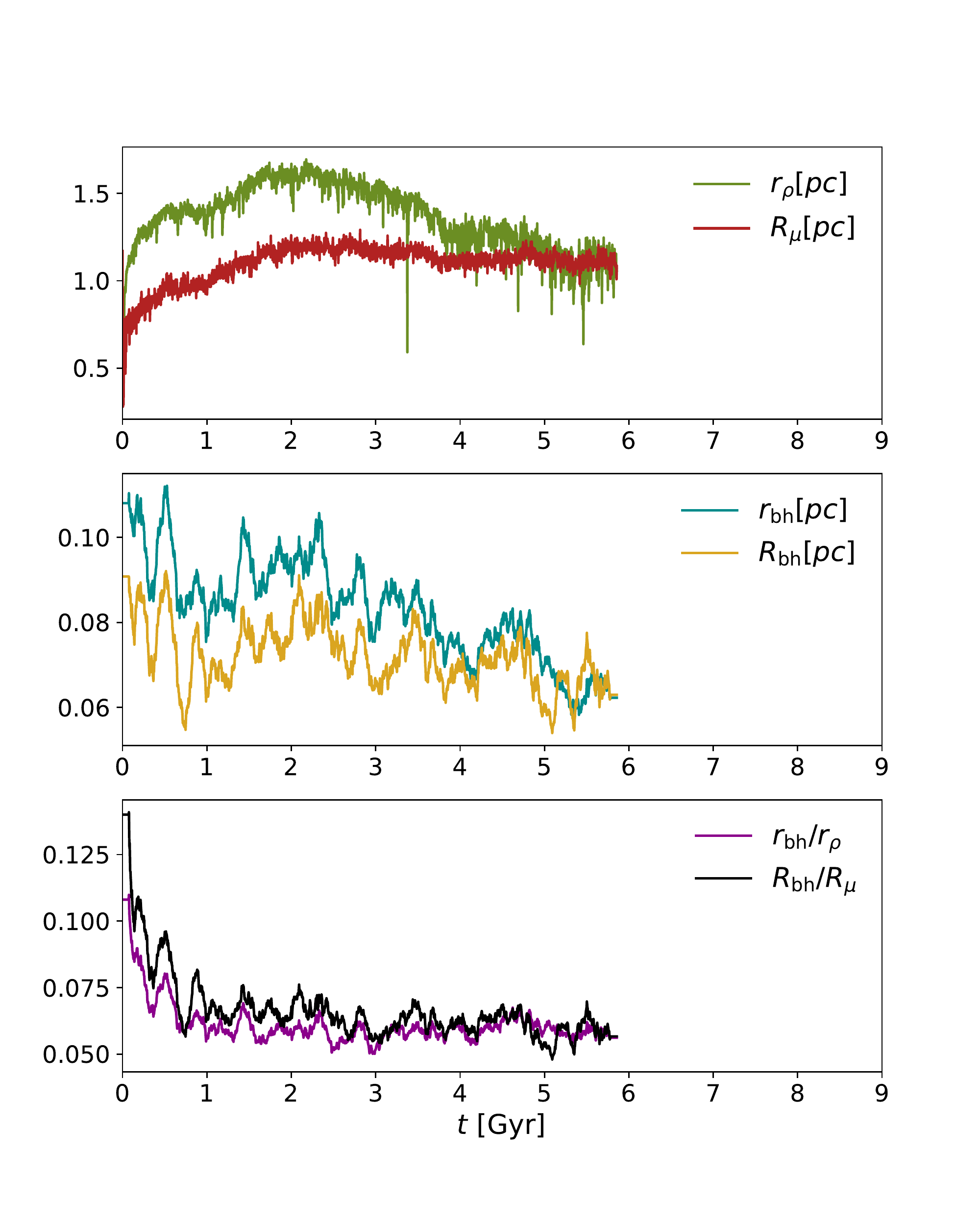}} &
\subfloat[]{\includegraphics[width=0.47\textwidth]{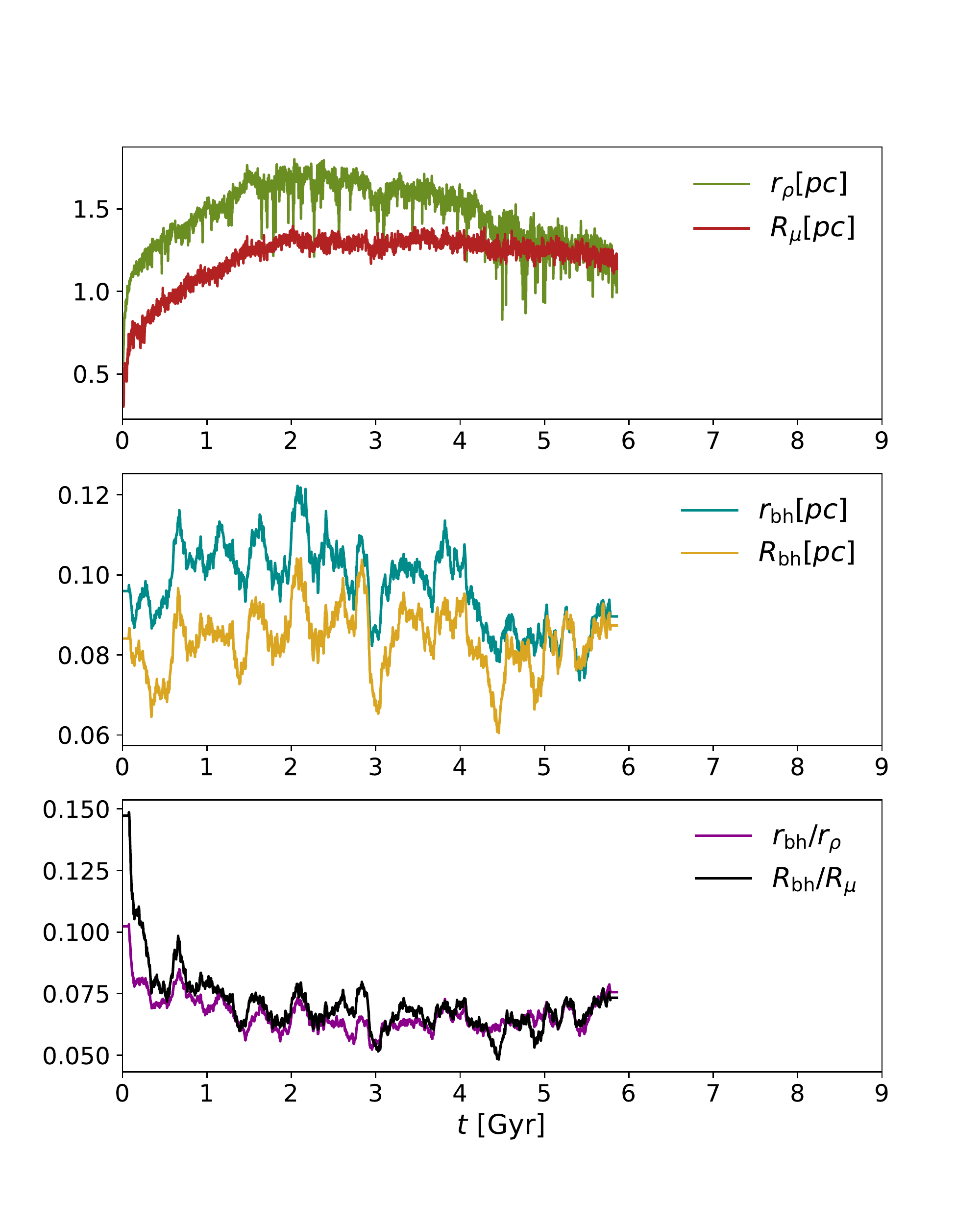}}
\\
\subfloat[]{\includegraphics[width=0.47\textwidth]{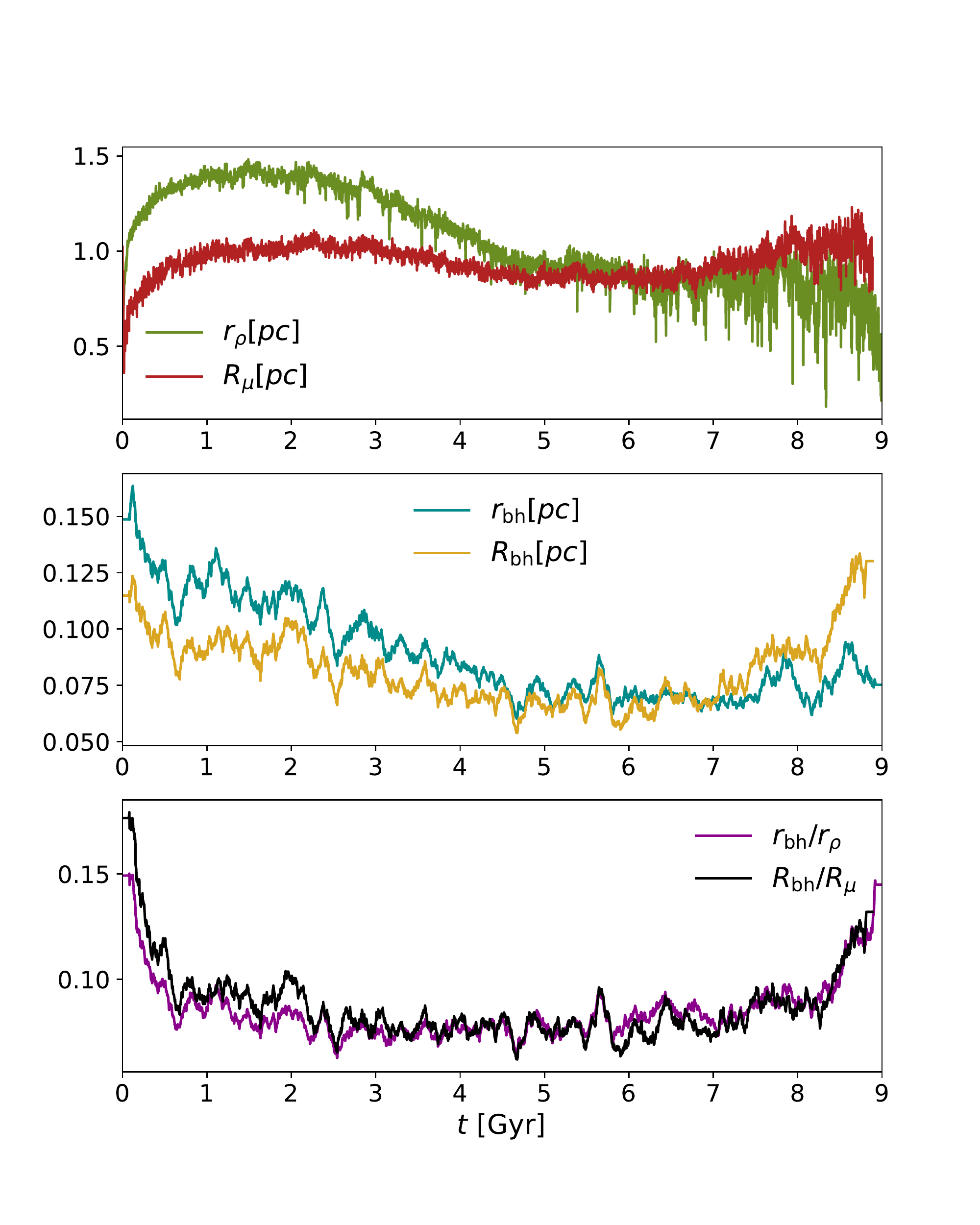}} &
\subfloat[]{\includegraphics[width=0.47\textwidth]{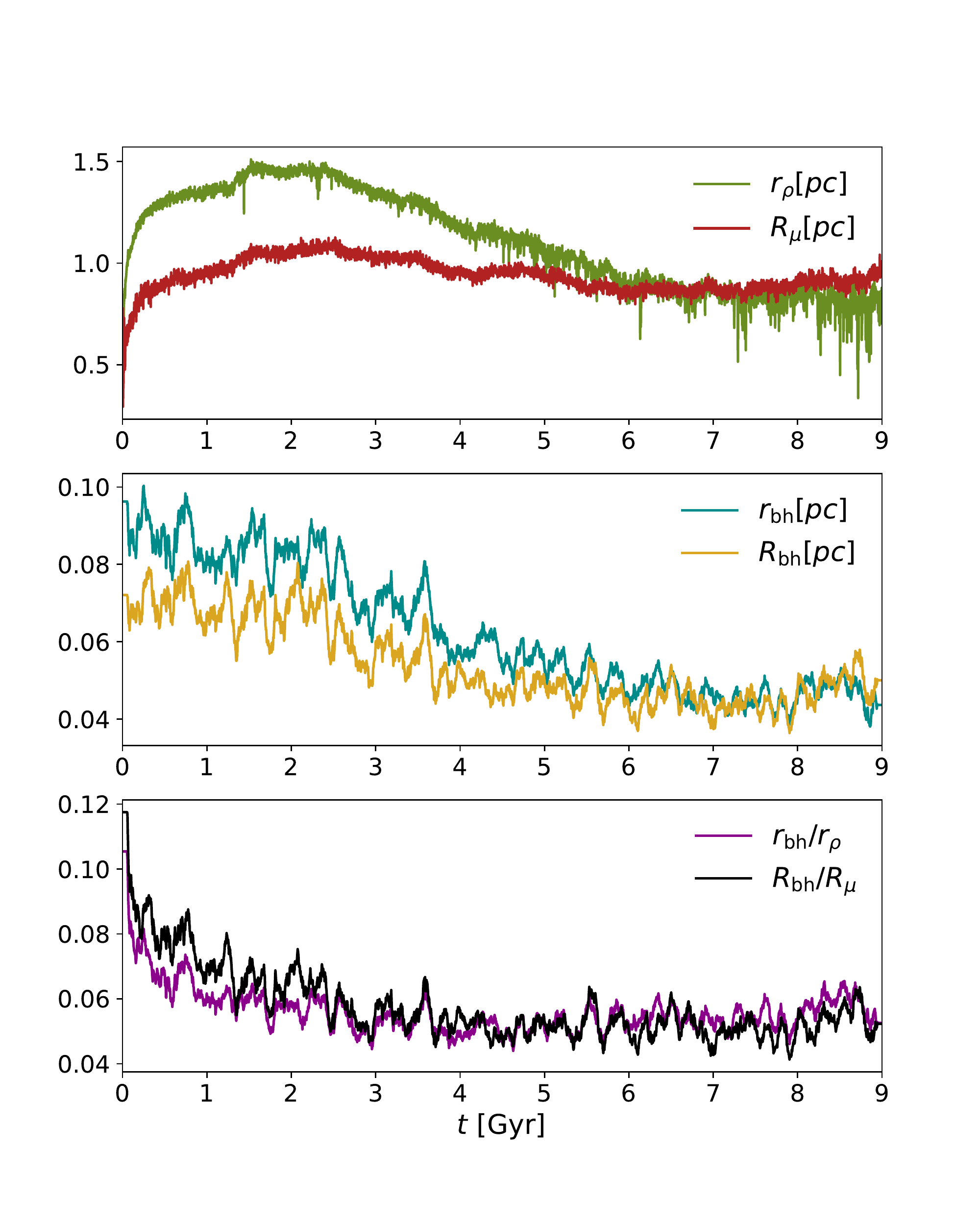}}
\\
\end{tabular}
\caption{Evolution of the 3D mass-weighted density radius $r_\rho$, the 2D luminosity-weighted surface brightness radius $R_\mu$, the intrinsic ($r_\mathrm{bh}$) and the projected ($R_\mathrm{bh}$) IMBH radial displacement for the 4 groups of simulations used in this paper (see Table \ref{tab:simul}).}
\label{fig:radii_evol}
\end{figure*}

In Fig.~\ref{fig:radii_evol} we plot the evolution of the 3D mass-weighted density radius $r_\rho$, of the 2D luminosity-weighted surface brightness radius $R_\mu$ and of the IMBH radial displacement, both intrinsic $r_\mathrm{bh}$ and projected $R_\mathrm{bh}$. In order to test the ability of the model in Equation (\ref{eq:displ}) to predict the IMBH displacement, we divide the range $4-5.5$ Gyr for simulations A-B and the range $4-7$ Gyr for simulations C-D into equal time intervals of 0.3 Gyr each. The time ranges are selected both to exclude the post core-collapse phase in which the density radius oscillates rapidly and to avoid the first part of the evolution dominated by stellar evolution and characterised by high discrepancy between the luminosity-based and mass-based radii. For each time interval we compare the direct measurement of the average IMBH displacement (presented in Subsection \ref{subsec:displ_sim}) with the displacement predicted by the model. We carry out this analysis for both 3D mass-weighted and 2D luminosity-weighted quantities. Accordingly, we replace the core terms in Equation (\ref{eq:displ}) with corresponding quantities in each case (see subsections below).

In both cases we are able to obtain a total of $N=100$ measurements for the relevant quantities in our model averaged over the $0.3$ Gyr time interval (the average over time is indicated using the brackets $\left<... \right>$). Then, the fit of the model to the data is carried out by maximising the Gaussian likelihood function 
\begin{equation}
\ln  \mathcal{L} = - \frac{1}{2} \sum_{i=1}^{N} 
\left[
\frac{f_i(A,\alpha,\beta)- \left<\log{(r_\mathrm{bh}/r_\mathrm{c})}\right>_i}{\delta_i} 
\right]^2,
\end{equation}
with respect to the free parameters $A$, $\alpha$ and $\beta$. Here $\delta_i$ is the standard error of $\left<\log{(r_\mathrm{bh}/r_\mathrm{c})}\right>_i$ and $f_i$ is the model prediction (see Equation \ref{eq:displ}) defined by
\begin{equation}
f_i \equiv \log{A} + 
\alpha \left<\log{\left( \frac{m_\mathrm{c}}{M_\mathrm{bh}} \right)}\right>_i + 
\beta \left<\log{\left( \frac{\sigma_\mathrm{c}^2 r_\mathrm{c}}{GM_\mathrm{c}} \right)}\right>_i.
\end{equation}
The maximisation process is performed using a Markov chain Monte Carlo (MCMC) estimator (see \citealt{foreman:13}).

\subsubsection{Mass-based analysis}
\label{subsec:mass_based_analysis}
 
For the mass-based analysis we rely on the maximum information available from the simulations, and use the intrinsic core radius with the intrinsic density radius defined by Equation (\ref{rd}) as proxy for $r_\mathrm{c}$. Similarly, the average stellar mass in the core $m_\mathrm{c}$ and the total core mass $M_\mathrm{c}$ are replaced by the same quantities calculated within the density radius (we will use the the subscript $\rho$ instead of $c$ to indicate quantities calculated within the density radius). Finally, we replace the core velocity dispersion $\sigma_\mathrm{c}$ with the 3D velocity dispersion $\sigma_\rho$, calculated from the velocity standard deviations along each axis of the stars (including dark stellar remnants) within the density radius.

\begin{figure*}
\begin{tabular}{cc}
\subfloat[]{\includegraphics[width=0.49\textwidth]{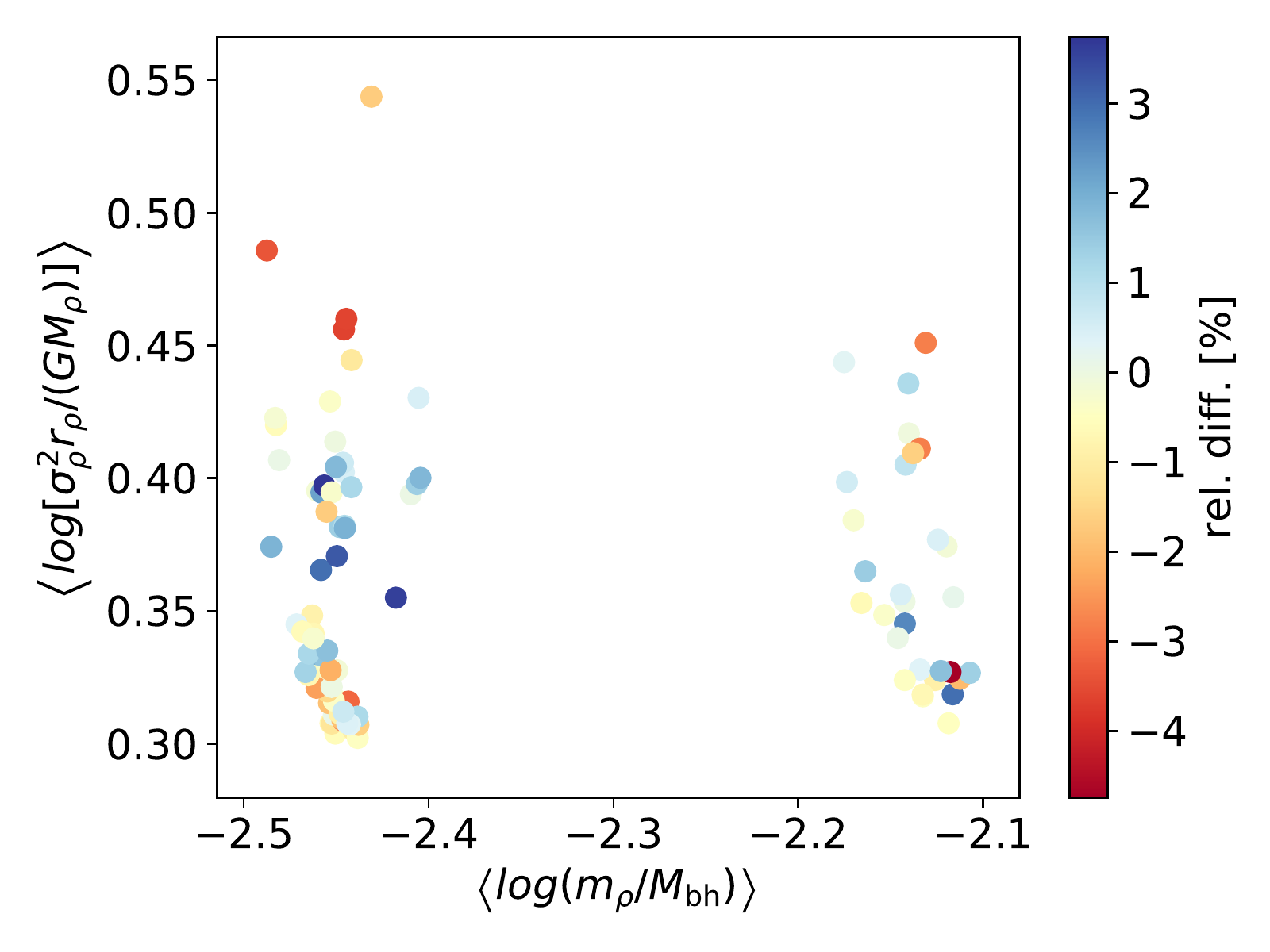}} &
\subfloat[]{\includegraphics[width=0.49\textwidth]{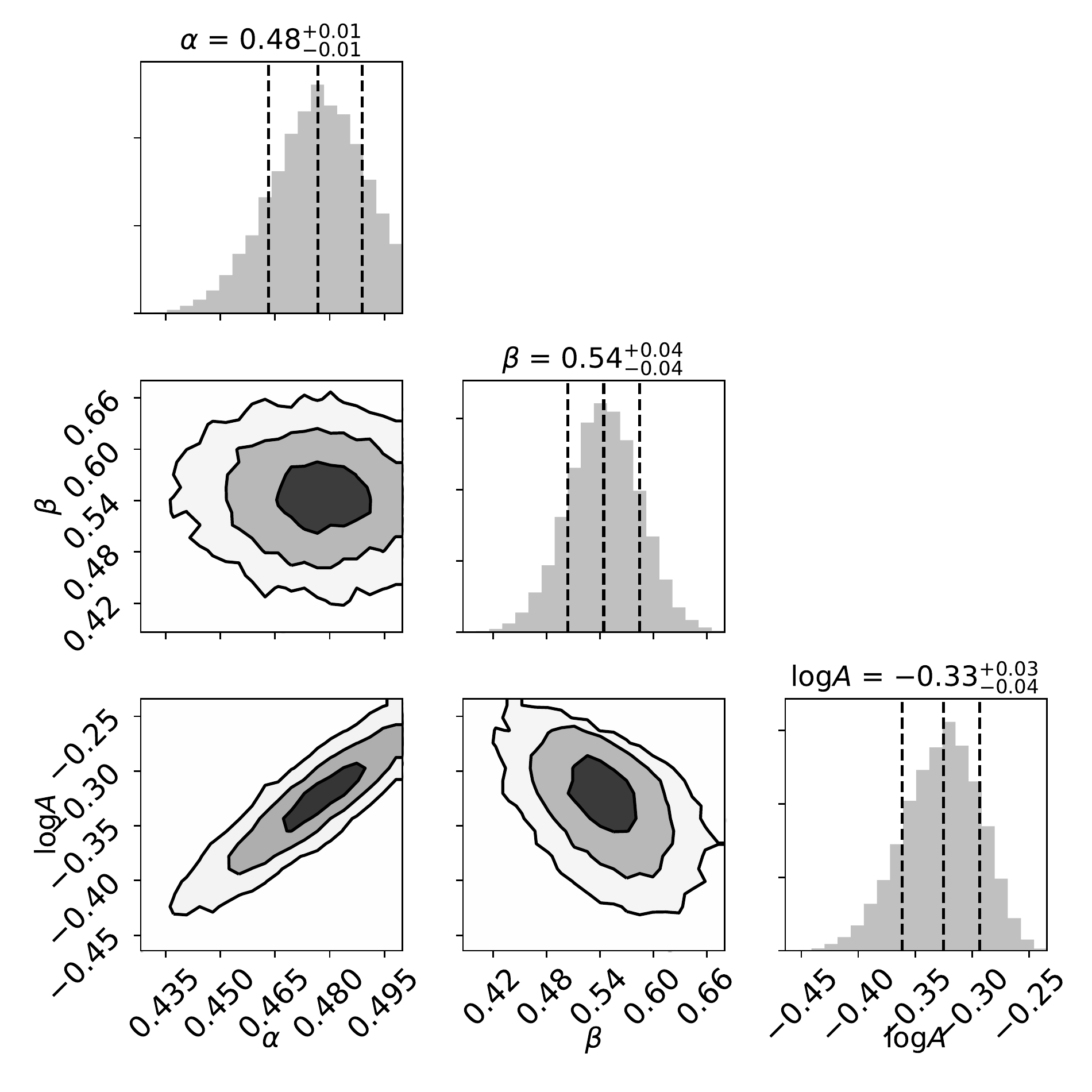}}
\end{tabular}
\caption
{ 
Best fit model for the IMBH displacement. Panel (a) shows a map of the IMBH displacement residuals, namely the relative difference between the displacement calculated with the model and the displacement measured in the simulation. Panel (b) shows the 1-D and 2-D histograms produced with the maximum likelihood estimator. The best fit values for $\log{A}$, $\alpha$ and $\beta$ are reported together with 1$\sigma$, 2$\sigma$ and 3$\sigma$ confidence levels. The chi-square per degree of freedom for this data set is $\tilde{\chi}_\mathrm{bf}^2=1.08$ (0.55$\sigma$).
}
\label{fig:best_fit_3d}
\end{figure*}

In Fig.~\ref{fig:best_fit_3d} we show the result of the fit. In particular, we plot the relative difference between the IMBH displacement (relative to the density radius) measured in the simulations to the corresponding displacement calculated with the best fit model. In addition, we report the histograms produced by the MCMC code to sample the parameter space. The best-fit value $\alpha=0.48\pm0.01$ suggests that the IMBH is very close to a state of complete energy equipartition with the surrounding stars. \footnote{Note that the near-complete energy equipartition for a single massive remnant does not imply that the whole cluster's core is in the same dynamical state, and our finding is consistent with the results of \citet{trenti:13}, which highlight that massive (dark) remnants have a higher degree of equipartition compared to visible stars because of their rarity.} Moreover, we find that a non-zero value of $\beta$ is needed in order to reproduce the data. This result supports the introduction of an additional parameter in our model related to the shape of the overall gravitational potential (see discussion in Subsection \ref{subsec:model}).  The chi-square per degree of freedom calculated with the best fit parameters is $\tilde{\chi}_\mathrm{bf}^2=1.08$, which corresponds to a $0.55\sigma$ deviation from the expected median value $\tilde{\chi}_\mathrm{bf}^2=1$, and indicates that our model describes the numerical data well.  

Finally, we test the quality of the fit performed with the model in Equation (\ref{eq:displ}) against other three models that rely on a lower number of free parameters. These models are readily obtained from Equation (\ref{eq:displ}) by imposing respectively (i) $\alpha=0.5$ and $\beta=0$; (ii) $\beta=0$ and (iii) $\alpha=0.5$. The first case represents the one-dimensional model of \cite{bahcall:76} expressed in Equation (\ref{eq:en_eq}), while the second case is the corresponding 2-dimensional version in which the hypothesis of complete energy equipartition has been relaxed. Finally, in the third case, the dynamical state of the core (as measured by $\beta$) is the only physical parameter, with the degree of energy equipartition fixed to its maximum value. 

\begin{figure*}
\includegraphics[width=0.49\textwidth]{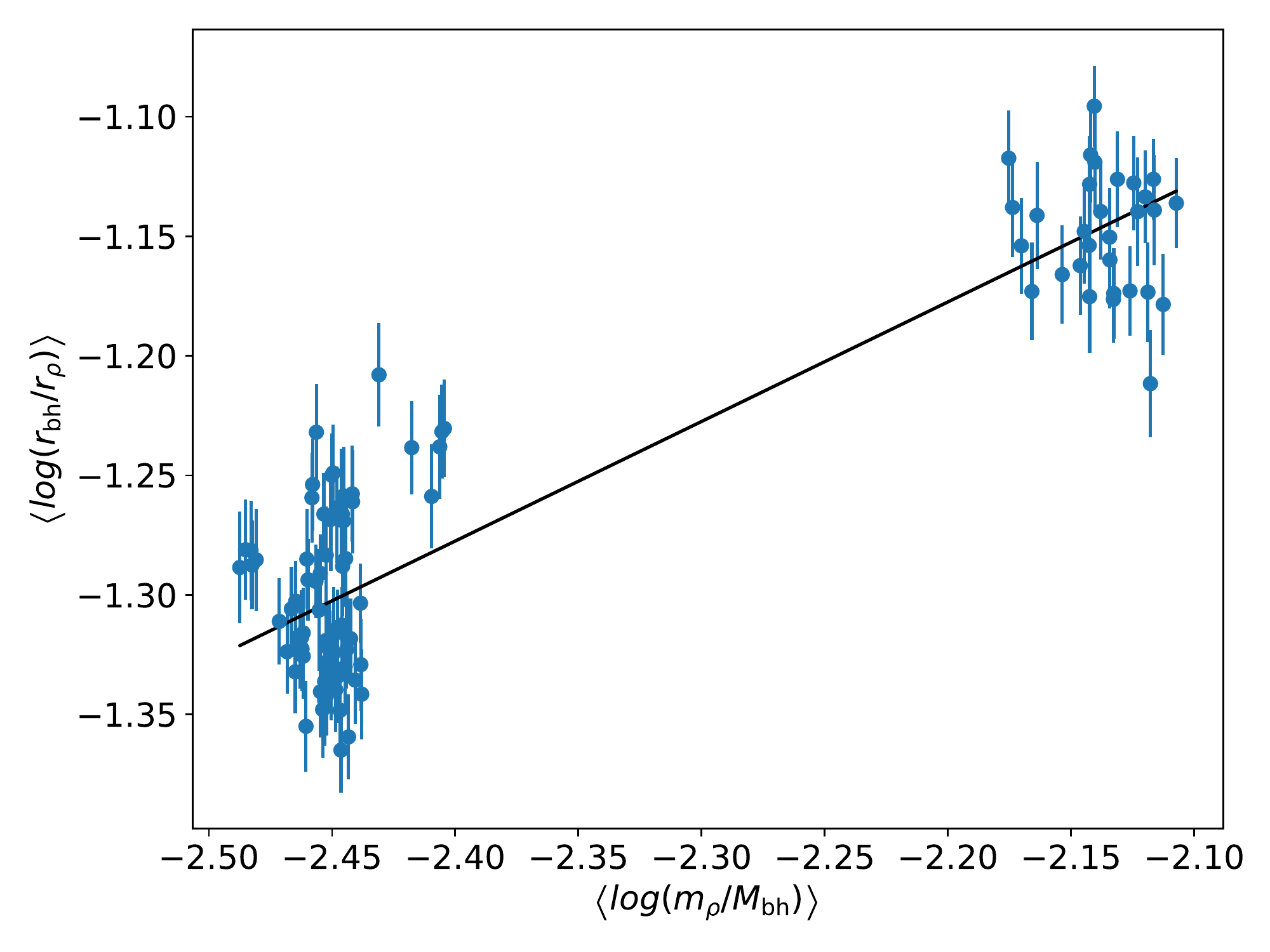}
\caption
{ 
Linear fit of the IMBH displacement predicted by Equation (\ref{eq:en_eq}), namely the model of Equation (\ref{eq:displ}) with constrains $\alpha=0.5$ and $\beta=0$. The mean relative displacement is reported as function of the ratio $m_\rho/M_\mathrm{bh}$ and the best linear fit of the data is shown. The chi-square per degree of freedom for this data set is $\tilde{\chi}_\mathrm{bf}^2=2.89$ (13.3$\sigma$).
}
\label{fig:Merritt_fit}
\end{figure*}

In Fig.~\ref{fig:Merritt_fit}, we plot the best-fit of the data through the most simple model of Equation (\ref{eq:en_eq}). The relevant best-fit values relative to the four different models are reported in Table \ref{tab:fits}. From the Akaike information criterion \footnote{The AIC is defined by $\text{AIC}=-2\ln \mathcal{L}_\mathrm{bf}+2k+2k(k+1)/(N-k-1)$, where $ \mathcal{L}_\mathrm{bf}$ is the maximum likelihood from the fit of a model with $k$ degrees of freedom to $N$ data points. The best model is the one which minimises AIC.} (see, e.g., \citealt{liddle:07}) we can conclude that the model in Equation (\ref{eq:displ}) (for which $\text{AIC} \approx 110$) and the analogous  model with the constraint $\alpha=0.5$ (for which $\text{AIC}\approx111$) are generally more appropriate to describe the IMBH dynamical behaviour when compared to the others (for which $\text{AIC}\approx290$). A likelihood ratio test for the model with $\alpha=0.5$ gives a value of $2.79$, indicating marginal significance (at the $90\%$ confidence level) that the modeling needs to allow for a departure from full-energy equipartition of the IMBH. Motivated by this, for the following analysis we adopt the most general model expressed by Equation (\ref{eq:displ}). However, we would expect to find similar results for the $\alpha=0.5$ model.

\begin{table*}
\centering
\caption{Quality of the fit. For four models (specified by Equation \ref{eq:displ}) with different degrees of freedom ($k$), we report the best-fit parameters ($\alpha_\mathrm{bf}$, $\beta_\mathrm{bf}$), the minimum chi-square ($\chi^2_\mathrm{bf}$),  the reduced chi-square ($\tilde{\chi}^2_\mathrm{bf}$), the deviation of $\tilde{\chi}^2_\mathrm{bf}$ from 1 in terms of the variance ($\sigma^2$) of the $\tilde{\chi}^2$ distribution, and the $\text{AIC}$ value.} 
\label{tab:fits}
\begin{tabular}{lccccccc}
\hline
$k$		&$\alpha_\mathrm{bf}$ 		&$\beta_\mathrm{bf}$		& $\chi^2_\mathrm{bf}$ 		& $\tilde{\chi}^2_\mathrm{bf}$ 		& $(\tilde{\chi}^2_\mathrm{bf}-1)/\sigma$ 		&$\text{AIC}$   \\
\hline

1	&0.5					&0	 			& 286.79	&2.89 	& 13.34	&288.80\\
2 	&$0.49\pm0.03$		&0	 			& 285.57	&2.91	& 13.40 	&289.57\\
2	&$0.5$ 				&$0.54\pm0.04$	& 107.36	&1.09	& 0.67 	&111.36\\
3	&$0.48\pm0.01$ 		&$0.55\pm0.04$ 	& 104.57 	&1.07	& 0.54 	&110.57\\
\hline
\end{tabular}
 \end{table*}
 
\renewcommand{\thesubfigure}{\alph{subfigure}}

\subsubsection{Luminosity-based analysis}

For an easier application of our model for the IMBH displacement to observed globular clusters, we carry out also an analysis in which the quantities involved are projected and luminosity-weighted. Thus we decide to replace the intrinsic core radius in the model presented in Subsection \ref{subsec:model}  with the projected surface-brightness density radius $R_\mu$, defined in Equation (\ref{Rmu}). Following the convention used for the mass-weighted analysis we use the subscript $\mu$ to indicate quantities calculated within $R_\mu$. With this convention, $m_\mu$ and $M_\mu$ represent the average stellar mass and the total mass calculated by considering all the stars (including dark remnants) enclosed in a projected circle with radius $R_\mu$. Finally, we indicate with $\sigma_{0,z}$ the standard deviation of the velocities along the z-axis of the luminous stars (namely main sequence stars with mass greater than $0.4M_\odot$) within a small circle around the centre (with radius $\approx 5\%$ of $R_\mu$). 

\begin{figure*}
\begin{tabular}{cc}
\subfloat[]{\includegraphics[width=0.49\textwidth]{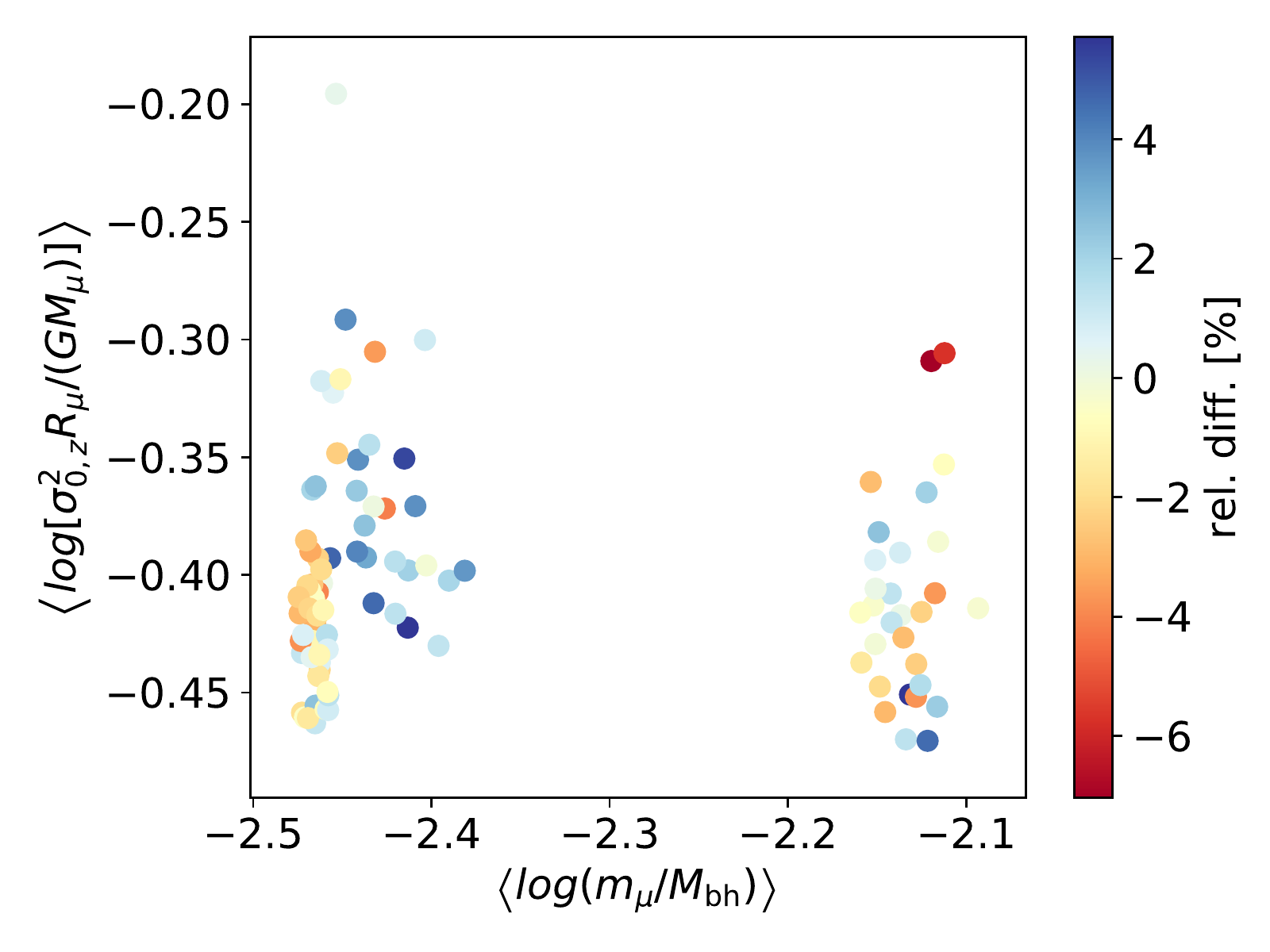}} &
\subfloat[]{\includegraphics[width=0.49\textwidth]{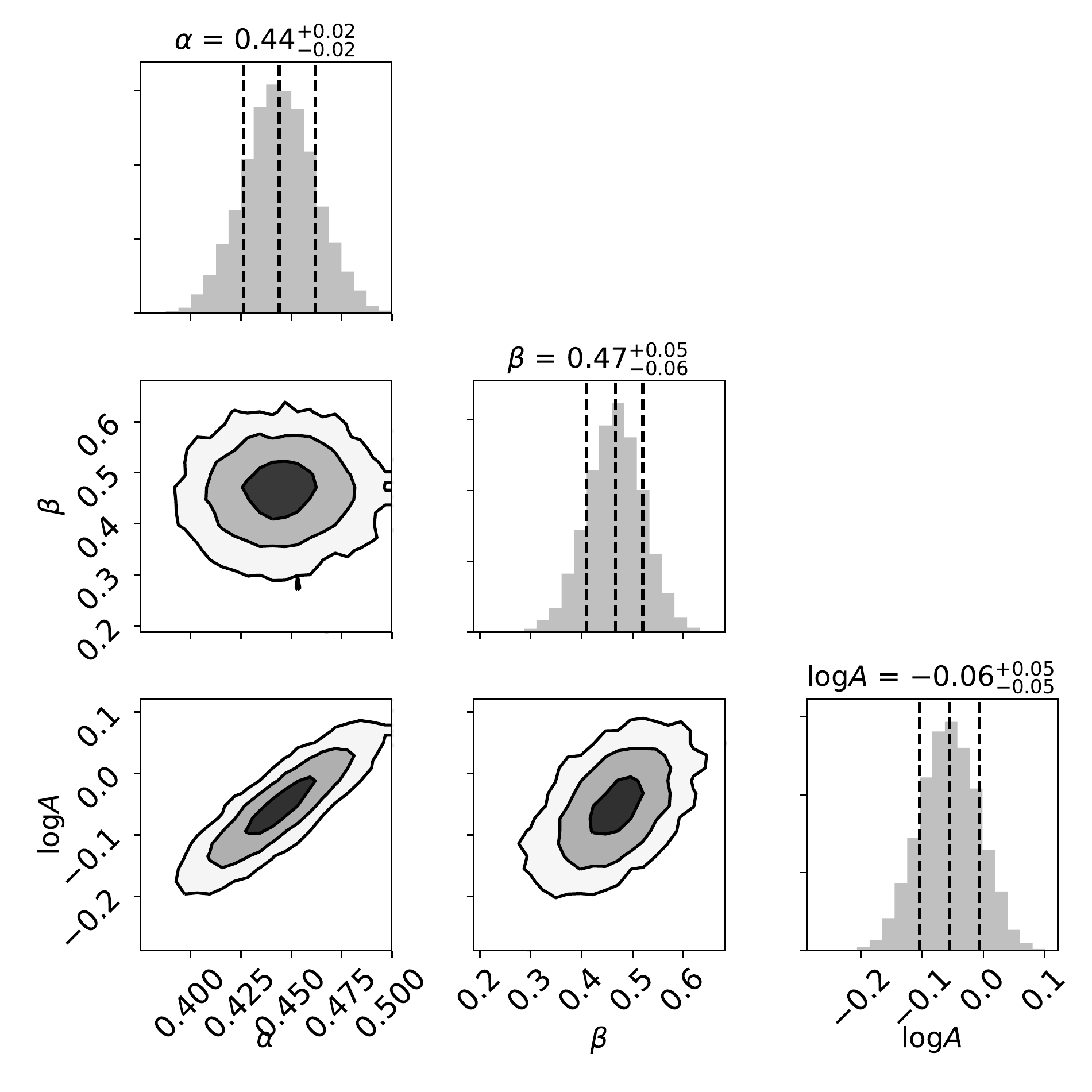}}
\end{tabular}
\caption
{ 
See Fig.~\ref{fig:best_fit_3d} for a detailed description. The chi-square per degree of freedom for this data set is $\tilde{\chi}_\mathrm{bf}^2=1.66$ (4.60$\sigma$).
}
\label{fig:best_fit_2d}
\end{figure*}

In Fig.~\ref{fig:best_fit_2d} we show the result of the fit based on luminosity-weighted and projected quantities. The chi-square per degree of freedom calculated with the best fit parameters is $\tilde{\chi}_\mathrm{bf}^2=1.66$, which corresponds to a $4.60\sigma$ deviation from the expected median value $\tilde{\chi}_\mathrm{bf}^2=1$. Overall, even if luminosity-based quantities are less effective as input for the dynamical modeling compared to mass-based measurements, we find good agreement between the best-fit parameters. This suggests that the model we present provides a basic yet effective tool to estimate the IMBH radial displacement on the basis of few parameters which are broadly available from GC observations. 

\renewcommand{\thesubfigure}{\alph{subfigure}}


\section{Wandering of putative IMBHs in Galactic globular clusters}
\label{sec:real_clusters}

With the model presented in Subsection \ref{subsec:model}, we have a tool to calculate the displacement expected in Galactic globular clusters. If applied to Equation (\ref{eq:displ}), the best fit values found in the luminosity-based analysis give the final version of the expected mean IMBH displacement:
\begin{eqnarray}
\label{eq:best_model}
\left<R_\mathrm{bh}\right> \approx 0.055 \left(\frac{5 \text{km/s}}{\sigma_{0,z}}\right)^{0.94} \left(\frac{m_\mu}{0.6 M_\odot}\right)^{0.44} \times \nonumber\\
\times \left(\frac{150 M_\odot}{M_\mathrm{bh}}\right)^{0.44} \times \left(\frac{1\text{pc}}{R_\mu}\right)^{0.47} \times  \nonumber\\
\times \left(\frac{M_\mu}{3\times10^3 M_\odot}\right)^{0.47}.
\end{eqnarray}

We consider the \citet{mclaughlin:05} catalogue to analyse the best fit King models of 85 Galactic GCs. We used the tabulated $W_0$, core radius, and total inferred mass of the cluster to constraint a King model. Then, with the use of the limepy software developed by \citet{gieles:15}, we derived relevant quantities as the central projected velocity dispersion ($\sigma_{0,z}$), and the total mass enclosed in the core radius ($\sim M_\mu$). Finally we identify $R_\mu$ with the tabulated projected core radius, and we considered a fixed average stellar mass $m_\mu=0.65~M_\odot$ and a fixed ratio $M_\mathrm{tot}/M_\mathrm{bh}=10^3$ for every cluster. 

\begin{figure*}
\includegraphics[width=\textwidth]{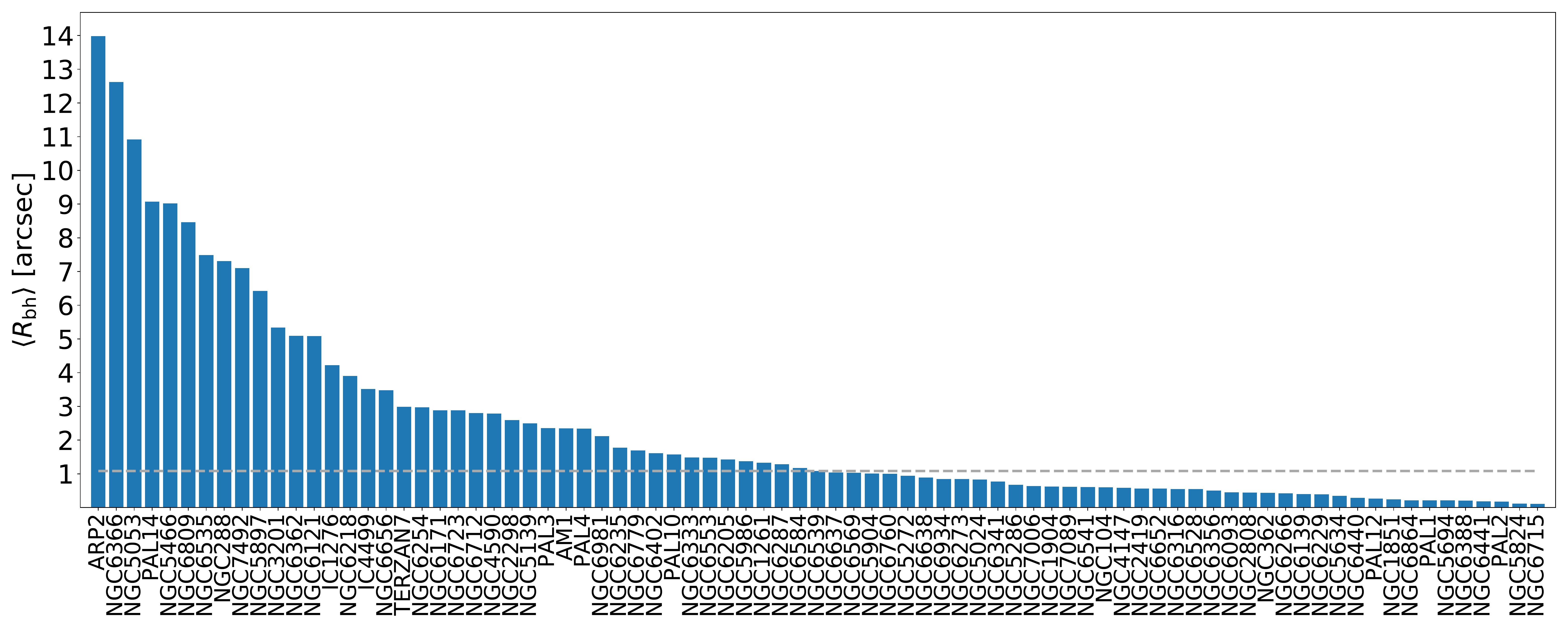} 
\caption
{ 
Expected IMBH radial displacement as calculated through Equation (\ref{eq:best_model}) with a fixed ratio $M_\mathrm{tot}/M_\mathrm{bh}=10^3$ for a sample of 85 Galactic GCs in the \citet{mclaughlin:05} catalogue. The dashed horizontal line represents a median value of $\approx1\arcsec$.
}
\label{fig:real_clusters}
\end{figure*}

In Fig.~\ref{fig:real_clusters} we plot the IMBH displacements calculated with our model for the selected sample of GCs. 
For the majority of the clusters the average IMBH radial displacement is around $1\arcsec$, with some outliers (NGC5053, NGC6366, and ARP2) showing a $\gtrsim10\arcsec$ displacement. According to our analysis, the debated case of $\omega$ Cen (see, e.g., the $\sim3.5\arcsec$ discrepancy in the determination of the centre position between \citealt{noyola:10} and \citealt{anderson:10}), is expected to show a rms displacement of $\approx 2.5\arcsec$ from the light center of the system. Another debated system in the literature is represented by NGC6388 (see \citealt{lanzoni:13} and \citealt{lutzgendorf:15}), which in our estimate shows a relatively modest rms displacement of less than $0.5\arcsec$. 

$\ $ 

In typical integrated-light integral field unit (IFU) observations the instrumental field of view is approximately $10\arcsec\times10\arcsec$ with a spaxel resolution of $0.3\arcsec-0.5\arcsec$, with the latter corresponding to the typical uncertainty in the centre determination (see, e.g., \citealt{lutzgendorf:13}). In this observational framework, the IMBH median displacement we estimate for Galactic GCs is not expected to introduce major systematic errors in the IMBH detection (see \citealt{devita:17}). However, effects of larger displacements ($\left<R_\mathrm{bh}\right>\gtrsim 2\arcsec$) would require tailored data-modeling comparison, since the effects of departure from spherical symmetry may affect the ability to correctly recover an unbiased BH mass. Furthermore, in the case of $\omega$ Cen, the best-fit mass inferred from spherical Jeans models may vary up to $~30\%$ when cluster centres with a $\sim10\arcsec$ separation are considered (see \citealt{noyola:10}). This implies that an accurate observational determination of the dynamical centre of the system and a modeling that account for wandering of a putative central IMBH would be crucial for a precise estimate of its mass and associated uncertainty. In particular, if an IMBH is present off-center and data are analysed through a standard spherically symmetric Jean model, then the recovered IMBH mass is expected to be under-estimated. 

\section{Discussion and conclusions} \label{sec:conclusions}

In this work we analysed the motion of IMBHs around the centres of globular clusters. First, we presented a simple model for the IMBH displacement (see Equation \ref{eq:displ}), which has been constructed based on few physical ingredients and on the comparison with realistic direct N-body simulations. The total number of free parameters of the model has been kept low in order to reduce the complexity of the dynamical processes involved in the IMBH random motion to two main aspects: the dynamical state of the core and the degree of energy equipartition between the IMBH and the field stars. A possible extension of this treatment might contemplate the inclusion of the effect that a IMBH companion should have on the binary's barycentre motion \citep{merritt:01}. However, for our simulations, we find that the IMBH radial displacement is not significantly affected by three-body scattering events, which we expect to represent a secondary aspect of the overall dynamics also in the case of more massive GCs ($\lesssim6\%$ relative correction). 

The negligible contribution of three-body scattering events to the IMBH displacement also implies that our results are likely to remain representative of more realistic simulations that include a non-zero fraction of primordial binaries. In fact, \cite{trenti:07} showed (see Section 6 in that paper) that while the presence of a central IMBH enhances the disruption rate of primordial binaries, its effect is indirect since binaries on orbits that would put them in at sufficiently close impact parameters have a low probability of reaching the IMBH without being disrupted first through three of four body encounters with other particles inside the sphere of influence of the BH. Thus, we expect that even in presence of primordial binaries the dominant energy exchanges with the IMBH would be through two-body encounters, with a modest overall enhancement of the typical displacement from the cluster center. 

After providing with a physical motivation for the model in Equation (\ref{eq:displ}), we focused on finding the set of model's parameters that best reproduces our numerical simulations. This analysis has been carried out for two distinct cases. In the first case, we considered three dimensional and mass-based quantities, getting advantage of the whole information available from the simulations. We find that the best fit model gives an overall good description of our data and generally offers a better performance when compared to related models with a lower number of free parameters (in particular, those models for which the dynamical state of the core is constrained a priori). The best-fit parameters indicate that the dynamical state of the core has to be considered in order to reproduce the data ($\beta\neq0$ in Equation \ref{eq:displ}), and that the IMBH is very close to a state of complete energy equipartition with the stars in the core ($\alpha\approx0.5$ in Equation \ref{eq:displ}). In the second case, we adopted projected and luminosity-weighted quantities in order to provide with a more direct tool for application to real observations. The results of the luminosity-based fit are consistent with the mass-based fit output, and are summarised by Equation (\ref{eq:best_model}), which gives the IMBH radial displacement as function of the IMBH mass and globular cluster structural parameters. 

We note that our modeling is focused on the long-term dynamical evolution of the simulated clusters (we limit the analysis to the time range $4-7$ Gyr), when massive stars have already evolved off the main sequence. For this reason we expect that our conclusions would not be critically altered by different choices for the IMF, the metallicity and the stellar evolutionary tracks adopted in the simulations, since these aspects affect primarily the early-time dynamical evolution of the simulated star clusters, whose memory is erased from the system over the relaxation timescale (below $1.5$ Gyr for a typical cluster). The only important exception is that, as shown in \cite{spera:15}, a different stellar evolution parameterisation would produce a different fraction of massive remnants. For example, more NSs and stellar-mass BHs would be present at late times as a consequence of a lower rate of mass loss from stellar winds, increasing the average mass in the core, and in turn the average IMBH displacement.

Finally, to illustrate an application of Equation (\ref{eq:best_model}), we resorted to the structural parameters catalog of \citet{mclaughlin:05} (which includes the majority of galactic objects) to derive the expected distribution of average IMBH radial displacements by assuming a fixed ratio for the total cluster mass to the IMBH mass. For the ratio $M_\mathrm{tot}/M_\mathrm{bh} = 10^3$, we find that the median value of the IMBH displacement is $\left< R_\mathrm{bh} \right> \approx1\arcsec$, with a few objects being significant outliers. In particular, predictions for Omega Centauri show an average offset from the centre $\left< R_\mathrm{bh} \right> \approx2.5\arcsec$. We note that given the lack of consensus on IMBH mass determinations in GCs, our assumptions rely on the uncertain extrapolation of the relations observed in galaxies between bulge and BH masses (e.g. see \citealt{gultekin:09}). However, they can be promptly rescaled to arbitrary BH masses through Equation (\ref{eq:best_model}), which is derived from an analytical modeling designed exactly to bypass the limitations of running a small number of N-body simulations that can explore only a limited mass range (in our case $M_\mathrm{tot}/M_\mathrm{bh} = 400-700$, which is within the $1\sigma$ uncertainty of the scaling relation derived by \citealt{lutzgendorf:13}). 

In conclusion, our findings suggest that while the median displacement is unlikely to significantly affect dynamical BH mass estimates, adding tailored dynamical modeling to include the IMBH displacement would lead to more precise estimates of both BH masses and associated systematic uncertainties. In particular, generalising spherical Jeans modeling to account for a separation between the center of the stars' gravitational potential and the center of the point-mass potential generated by an IMBH would be the most useful improvement. In this framework, higher orders in the multipole expansion of the combined gravitational potential should be included in the modeling process and would potentially help in solving the tension between different interpretations of velocity dispersion data for globular clusters in which a central IMBH has been claimed to be present.

\section*{Acknowledgements}

We thank the referee for helpful comments and suggestions. This work was partially supported by the A.A.H. Pierce Bequest at the University of Melbourne. M.M. acknowledges helpful conversations with S. Tremaine. M.M. is grateful for support provided by NASA through Einstein Postdoctoral Fellowship grant number PF6-170155 awarded by the Chandra X-ray Center, which is operated by the Smithsonian Astrophysical Observatory for NASA under contract NAS8- 03060. This research made use of Astropy, a community-developed core Python package for Astronomy \citep{astropy}.





\bibliographystyle{mnras}
\bibliography{biblio} 


\end{document}